\documentclass[longauth]{aa}
\usepackage[colorlinks=true, allcolors=blue]{hyperref}
\usepackage{graphicx}
\usepackage[varg]{txfonts}
\usepackage{multirow}
\usepackage{array}
\usepackage{silence}
\usepackage{ulem}
\usepackage{mathtools}
\usepackage{amsmath}
\usepackage{booktabs}
\usepackage{longtable}
\usepackage{float}
\usepackage{tabularx}
\usepackage{graphicx}
\usepackage{threeparttable}
\usepackage{tabularx}
\usepackage{multirow}
\usepackage{amsmath}
\usepackage{amssymb}
\usepackage{flushend}

\newcolumntype{Y}{>{\centering\arraybackslash}X}

\newcommand{\Ha} {\mbox{H$\alpha$}\,}

\newcommand{\Feii} {\ion{Fe}{ii}\,}

\newcommand{\Caii} {\ion{Ca}{ii}\,}

\newcommand{\Ciii} {\ion{C}{iii}\,}

\newcommand{\Hei} {\ion{He}{i}\,}
\newcommand{\Heii} {\ion{He}{ii}\,}

\newcommand{\Nii} {\ion{N}{ii}\,}
\newcommand{\Niii} {\ion{N}{iii}\,}
\newcommand{\Oi} {\ion{O}{i}\,}

\newcommand{\Mgi} {\ion{Mg}{i}\,}
\newcommand{\Mgii} {\ion{Mg}{ii}\,}

\newcommand{\msun}{\mbox{M$_{\odot}$\,}}

\newcommand{\kms}{\mbox{$\rm{km}\,s^{-1}$}}

\begin{document}

   \title{Massive stars exploding in a He-rich circumstellar medium}

   \subtitle{XII. SN\,2024acyl: A fast, linearly declining Type Ibn supernova with early flash-ionisation features}

   \authorrunning{Y.-Z. Cai et al.} 
   \titlerunning{SN\,2024acyl}
   
\author{
    Y.-Z. Cai  \inst{\ref{inst1},\ref{inst2},\ref{inst3}}\thanks{Corresponding authors: caiyongzhi@ynao.ac.cn (CYZ)} \and
    A. Pastorello \inst{\ref{inst3}} \and
    K. Maeda\inst{\ref{inst4}} \and
    J.-W.~Zhao\inst{\ref{inst5},\ref{inst6}} \and
    Z.-Y. Wang \inst{\ref{inst7},\ref{inst8}} \and
    Z.-H. Peng\inst{\ref{inst9}} \and
    A. Reguitti \inst{\ref{inst10},\ref{inst3}} \and
    L.~Tartaglia \inst{\ref{inst11}} \and
    A.~V.~Filippenko \inst{\ref{inst13}} \and
    Y. Pan\inst{\ref{inst5},\ref{inst6}} \and
    G. Valerin \inst{\ref{inst3}} \and
    B. Kumar\inst{\ref{inst5},\ref{inst6}} \and
    Z. Wang \inst{\ref{inst1},\ref{inst2},\ref{inst7}} \and
    M. Fraser \inst{\ref{inst20}} \and
    J.~P.~Anderson \inst{\ref{inst14},\ref{inst15}} \and
    S. Benetti \inst{\ref{inst3}} \and
    S.~Bose \inst{\ref{inst16}} \and
    T. G. Brink \inst{\ref{inst13}} \and
    E. Cappellaro\inst{\ref{inst3}} \and
    T.-W. Chen\inst{\ref{inst17}} \and
    X.-L. Chen\inst{\ref{inst5},\ref{inst6}} \and
    N.~Elias-Rosa\inst{\ref{inst3},\ref{inst19}} \and
    A. Esamdin \inst{\ref{inst25},\ref{inst7}}\and
    A.~Gal-Yam \inst{\ref{inst24}}\and
    M. Gonz\'alez-Ba\~nuelos\inst{\ref{inst19},\ref{inst21}} \and
    M. Gromadzki \inst{\ref{inst46}} \and
    C.~P.~Guti\'errez \inst{\ref{inst21},\ref{inst19}} \and
    C. Inserra\inst{\ref{inst22}}\and
    A. Iskandar \inst{\ref{inst25},\ref{inst7}}\and
    T.~Kangas\inst{\ref{inst26},\ref{inst23}} \and
    E. Kankare\inst{\ref{inst23}} \and
    T.~Kravtsov\inst{\ref{inst47}}\and
    H.~Kuncarayakti\inst{\ref{inst23},\ref{inst26}} \and
    L.-P.~Li\inst{\ref{inst1},\ref{inst2}} \and
    C.-X. Liu\inst{\ref{inst5},\ref{inst6}} \and
    X.-K. Liu\inst{\ref{inst5},\ref{inst6}} \and
    P.~Lundqvist\inst{\ref{inst18}} \and
    K.~Matilainen\inst{\ref{inst23},\ref{inst27}} \and
    S.~Mattila\inst{\ref{inst44},\ref{inst45}}\and
    S. Moran\inst{\ref{inst30}} \and
    T.~E.~M\"uller-Bravo\inst{\ref{inst28},\ref{inst29}} \and
    T. Nagao\inst{\ref{inst31},\ref{inst32},\ref{inst33}} \and
    T. Petrushevska\inst{\ref{inst38}} \and
    G. Pignata\inst{\ref{inst34}} \and
    I.~Salmaso\inst{\ref{inst35},\ref{inst3}} \and
    S. J. Smartt\inst{\ref{inst36},\ref{inst37}} \and
    J. Sollerman\inst{\ref{inst18}} \and
    S. Srivastav \inst{\ref{inst36},\ref{inst37}} \and
    M. D. Stritzinger \inst{\ref{inst16}} \and
    L.-T. Wang \inst{\ref{inst25}} \and
    S.-Y.~Yan  \inst{\ref{inst39}}\and
    Y.~Yang \inst{\ref{inst39}} \and
    Y.-P.~Yang\inst{\ref{inst5},\ref{inst6}} \and
    W. Zheng \inst{\ref{inst13}} \and
    X.-Z. Zou\inst{\ref{inst5},\ref{inst6}} \and
    L.-Y. Chen\inst{\ref{inst39}} \and
    X.-L. Du\inst{\ref{inst5},\ref{inst6}} \and
    Q.-L.~Fang\inst{\ref{inst40}} \and
    A.~Fiore \inst{\ref{inst11},\ref{inst3}} \and
    F.~Ragosta\inst{\ref{inst41},\ref{inst35}} \and
    S.~Zha\inst{\ref{inst1},\ref{inst2}} \and
    J.-J. Zhang\inst{\ref{inst1},\ref{inst2}} \and
    X.-W. Liu\inst{\ref{inst5},\ref{inst6}} \and
    J.-M. Bai\inst{\ref{inst1},\ref{inst2}} \and
    B.~Wang \inst{\ref{inst1},\ref{inst2}}\thanks{wangbo@ynao.ac.cn (WB)} \and
    X.-F.~Wang \inst{\ref{inst39},\ref{inst42}}\thanks{wang\_xf@mail.tsinghua.edu.cn (WXF)}
}

\institute{
    \label{inst1}Yunnan Observatories, Chinese Academy of Sciences, Kunming 650216, P.R. China\and
    \label{inst2}International Centre of Supernovae, Yunnan Key Laboratory, Kunming 650216, P.R. China\and 
    \label{inst3}INAF - Osservatorio Astronomico di Padova, Vicolo dell'Osservatorio 5, 35122 Padova, Italy \and
    \label{inst4}Department of Astronomy, Kyoto University, Kitashirakawa-Oiwake-cho, Sakyo-ku, Kyoto 606-8502, Japan  \and
    \label{inst5}South-Western Institute for Astronomy Research, Yunnan University, Kunming 650500, P.R. China  \and
    \label{inst6}Yunnan Key Laboratory of Survey Science, Yunnan University, Kunming, Yunnan 650500, P.R. China   \and
    \label{inst7}School of Astronomy and Space Science, University of Chinese Academy of Sciences, Beijing 100049, P.R. China   \and
    \label{inst8}National Astronomical Observatories, Chinese Academy of Sciences, Beijing 100101, P.R. China \and
    \label{inst9}School of Electronic Science and Engineering, Chongqing University of Posts and Telecommunications, Chongqing 400065, P.R. China    \and
    \label{inst10}INAF - Osservatorio Astronomico di Brera, Via E. Bianchi 46, 23807 Merate (LC), Italy    \and
    \label{inst11}INAF - Osservatorio Astronomico d'Abruzzo, Via M. Maggini snc, 64100 Teramo, Italy   \and   
    \label{inst13}Department of Astronomy, University of California, Berkeley, CA 94720-3411, USA    \and 
    \label{inst20}School of Physics, O'Brien Centre for Science North, University College Dublin, Belfield, Dublin 4, Ireland  \and
    \label{inst14}European Southern Observatory, Alonso de C\'{o}rdova 3107, Casilla 19, Santiago, Chile   \and
    \label{inst15}Millennium Institute of Astrophysics (MAS), Nuncio Monse\~{n}or S\`{o}tero Sanz 100, Providencia, Santiago, 8320000 Chile    \and
    \label{inst16}Department of Physics and Astronomy, Aarhus University, Ny Munkegade 120, DK-8000 Aarhus C, Denmark   \and
    \label{inst17}Graduate Institute of Astronomy, National Central University, 300 Jhongda Road, 32001 Jhongli, Taiwan   \and
    \label{inst19}Institute of Space Sciences (ICE, CSIC), Campus UAB, Carrer de Can Magrans, s/n, E-08193 Barcelona, Spain  \and
    \label{inst25}Xinjiang Astronomical Observatory, Chinese Academy of Sciences, Urumqi, Xinjiang, 830011, P.R. China   \and
    \label{inst24}Department of Particle Physics and Astrophysics, Weizmann Institute of Science, 76100 Rehovot, Israel  \and
    \label{inst21}Institut d'Estudis Espacials de Catalunya (IEEC), 08860 Castelldefels (Barcelona), Spain   \and
    \label{inst46}Astronomical Observatory, University of Warsaw, Al. Ujazdowskie 4, 00-478 Warszawa, Poland \and
    \label{inst22}Cardiff Hub for Astrophysics Research and Technology, School of Physics \& Astronomy, Cardiff University, Queens Buildings, The Parade, Cardiff, CF24 3AA, UK   \and
    \label{inst26}Finnish Centre for Astronomy with ESO (FINCA), Quantum, Vesilinnantie 5, University of Turku, FI-20014 Turku, Finland \and
    \label{inst23}Tuorla Observatory, Department of Physics and Astronomy, University of Turku, FI-20014 Turku, Finland    \and
    \label{inst47}Department of Physics \& Astronomy, University of Turku, Vesilinnantie 5, Turku, FI-20014, Finland \and
    \label{inst18}The Oskar Klein Centre, Department of Astronomy, Stockholm University, AlbaNova, SE-10691 Stockholm, Sweden  \and
    \label{inst27}Nordic Optical Telescope, Aarhus Universitet, Rambla Jos\'e Ana Fern\'andez P\'erez 7, local 5, E-38711 San Antonio, Bre\~na Baja, Santa Cruz de Tenerife, Spain   \and
    \label{inst44}Department of Physics and Astronomy, University of Turku, FI-20014 Turku,  Finland  \and
    \label{inst45}School of Sciences, European University Cyprus, Diogenes Street, Engomi, 1516 Nicosia, Cyprus  \and
    \label{inst30}School of Physics and Astronomy, University of Leicester, University Road, Leicester LE1 7RH, UK   \and
    \label{inst28}School of Physics, Trinity College Dublin, The University of Dublin, Dublin 2, Ireland   \and
    \label{inst29}Instituto de Ciencias Exactas y Naturales (ICEN), Universidad Arturo Prat, Iquique, Chile \and
    \label{inst31}Department of Physics and Astronomy, University of Turku, FI-20014 Turku, Finland   \and
    \label{inst32}Aalto University Mets\"ahovi Radio Observatory, Mets\"ahovintie 114, 02540 Kylm\"al\"a, Finland \and
    \label{inst33}Aalto University Department of Electronics and Nanoengineering, P.O. BOX 15500, FI-00076 AALTO, Finland   \and
    \label{inst38}Center for Astrophysics and Cosmology, University of Nova Gorica, Vipavska 11c, 5270 Ajdov\v{s}\v{c}ina, Slovenia     \and
    \label{inst34}Instituto de Alta Investigaci\'on, Universidad de Tarapac\'a, Casilla 7D, Arica, Chile   \and
    \label{inst35}INAF - Osservatorio Astronomico di Capodimonte, Salita Moiariello 16, 80131 Napoli, Italy    \and
    \label{inst36}Department of Physics, University of Oxford, Denys Wilkinson Building, Keble Road, Oxford OX1 3RH, UK    \and
    \label{inst37}Astrophysics Research Centre, School of Mathematics and Physics, Queen’s University Belfast, Belfast BT7 1NN, UK     \and
    \label{inst39}Department of Physics, Tsinghua University, Beijing 100084, P.R. China     \and
    \label{inst40}National Astronomical Observatory of Japan, National Institutes of Natural Sciences, 2-21-1 Osawa, Mitaka, Tokyo 181-8588, Japan    \and
    \label{inst41}Dipartimento di Fisica “Ettore Pancini”, Università di Napoli Federico II, Via Cinthia 9, 80126 Naples, Italy \and
    \label{inst42}Purple Mountain Observatory, Chinese Academy of Sciences, Nanjing, 210023, P.R. China  
}

   \date{Received November 1, 2025; accepted January, 22 2026}
 
  \abstract
  {
    We present a photometric and spectroscopic analyses of the Type Ibn supernova (SN) 2024acyl. It rises to an absolute magnitude peak of $M_{o} = -17.58 \pm 0.15 \, \mathrm{mag}$ in 10.6 days, and displays a rapid linear post-peak light-curve decline in all bands (e.g. $\gamma_{0-60}($V$) = 0.097 \pm 0.002$ $\mathrm{mag \, day^{-1}}$), similar to most SNe Ibn. The optical pseudobolometric light curve peaks at ($3.5\pm0.8) \times 10^{42}$ erg s$^{-1}$, with a total radiated energy of $(5.0\pm0.4) \times 10^{48}$ erg. The spectra are dominated by a blue continuum at early stages, with narrow P-Cygni \Hei~lines and flash-ionisation emission lines of C {\sc iii}, N {\sc iii}, and He {\sc ii}. The P-Cygni \Hei~features gradually evolve and become emission-dominated in late-time spectra.  The \Ha~line is detected throughout the entire spectral evolution, which indicates that the circumstellar material (CSM) is helium-rich with some residual amount of hydrogen. Our multi-band light-curve modelling yields estimates of the ejecta mass of $M_{\mathrm{ej}}$ = 
    $0.49^{+0.11}_{-0.09} \, \msun$ with a kinetic energy of $E_{\mathrm{k}} = 0.06^{+0.01}_{-0.01} \times 10^{51}\,\mathrm{erg}$, and a $^{56}\mathrm{Ni}$ mass of $M_{\mathrm{Ni}} = 0.018 \, \msun$. The inferred CSM properties are characterised by a mass of $M_{\rm{CSM}} = 0.51^{+0.05}_{-0.04}$ \msun, an inner radius of $R_0$=$17.8^{+3.6}_{-3.0}$ AU, and a density of $\rho_{\mathrm{CSM}} = (8.3_{-1.2}^{+2.7})\times10^{-12} \, \mathrm{g\,cm^{-3}}$. 
    The multi-epoch spectra are well reproduced by the CMFGEN/ \texttt{he4p0} model, corresponding to a He-ZAMS mass of 4~M$_\odot$ (H-ZAMS mass 18.11~M$_\odot$, pre-SN mass 3.16~M$_\odot$). These findings are consistent with a scenario of an SN powered by ejecta-CSM interaction originating from a low-mass helium star that evolved within an interacting binary system where the CSM with some residual hydrogen may originate from the mass-transfer process. We also discuss an extreme scenario involving the possible merger of a helium white dwarf. 
    In addition, a channel of core-collapse explosion of a late-type Wolf-Rayet (WR) star with hydrogen, or a transitional star between an Of and a WR type (e.g. an Ofpe/WN9 star) with fallback accretion cannot be entirely ruled out.}

   \keywords{circumstellar matter -- supernovae: general -- supernovae: individual: SN\,2024acyl} 

   \maketitle

\nolinenumbers
\section{Introduction}

Type~Ibn supernovae (SNe~Ibn) represent a subtype of stellar explosions distinguished by relatively narrow ($\sim$1000 km s$^{-1}$) helium emission lines and weak (or no) evidence of hydrogen lines in their spectra, suggesting the presence of He-rich circumstellar material \citep[CSM;][]{Smith2017hsn..book..403S, Gal-Yam2017hsn..book..195G}. SN~1999cq was the first SN~Ibn identified with typical Type Ib spectral features superimposed with the narrow \Hei~lines \citep{Matheson2000AJ....119.2303M}; however, the formal designation of this new SN type was introduced by \citet{2008MNRAS.389..113P}, after the study of the prototypical Type Ibn SN~2006jc \citep[e.g.][]{Foley2007ApJ...657L.105F, Pastorello2007Natur.447..829P, Anupama2009MNRAS.392..894A}.  This class is defined by analogy with SNe IIn, which show relatively narrow H lines with full width half maximum intensity (FWHM) velocities ranging from a few hundred to $\sim 1000$ \kms, arising from the interaction of SN ejecta with the surrounding dense H-rich CSM \citep[][]{Schlegel1990MNRAS.244..269S,Filippenko1997ARA&A..35..309F,Fraser2020RSOS....700467F}.

A few SNe Ibn have displayed transitional spectra between classical Type Ibn and IIn SNe, with H and He~I lines having comparable strengths. This small sample of Type Ibn and IIn events includes SNe 2005la \citep{Pastorello2008MNRAS.389..131P}, 2010al \citep{Pastorello2015MNRAS.449.1921P}, 2011hw \citep{Smith2012MNRAS.426.1905S, Pastorello2015MNRAS.449.1921P}, 2020bqj \citep{Kool2021A&A...652A.136K}, and 2021foa \citep{Reguitti2022A&A...662L..10R,Farias2024ApJ...977..152F,Gangopadhyay2025MNRAS.537.2898G}.  Their moderately rich CSM suggests a continuity in properties between SNe Ibn and SNe IIn \citep[][]{Smith2012MNRAS.426.1905S, Pastorello2015MNRAS.449.1921P, Reguitti2022A&A...662L..10R}. 

Other markers indicating the presence of CSM are the very short duration ($\leq$ 10 days), narrow high-ionisation emission lines detected in very young SNe of various types \citep[e.g.][]{Gal-Yam2014Natur.509..471G,Shivvers2015ApJ...806..213S,Khazov2016ApJ...818....3K,Yaron2017NatPh..13..510Y,Zhang2023SciBu..68.2548Z,Zhang2024ApJ...970L..18Z,Bostroem2023ApJ...956L...5B}. These features are directly related to the effects of the shock breakout and arise from the recombination of the flash-ionised CSM \citep[e.g.][]{Gal-Yam2014Natur.509..471G,Yaron2017NatPh..13..510Y} and the interaction with a dilute wind inside the dense shell \citep{Fransson1996ApJ...461..993F}. To date, only a handful of Type Ibn events have occasionally been observed with flash signatures, including SNe~2010al \citep{Pastorello2015MNRAS.449.1921P}, 2019cj \citep{Wang2024A&A...691A.156W}, 2019uo \citep{Gangopadhyay2020ApJ...889..170G}, 2019wep \citep{Gangopadhyay2022ApJ...930..127G}, and 2023emq \citep{Pursiainen2023ApJ...959L..10P}.

Although SNe~Ibn exhibit some diversity in their spectra, they typically display an overall photometric homogeneity \citep[see e.g.][]{Pastorello2016MNRAS.456..853P, Hosseinzadeh2017ApJ...836..158H, Wang2025A&A...700A.156W, Dong2025arXiv251103926D}. The light curves of SNe~Ibn typically exhibit fast rise times ($\sim$7 days), rapid post-peak declines (0.05--0.15~mag~$\rm{day^{-1}}$), and luminous peak absolute magnitudes ($M \approx -19$~mag). However, several outliers exist, such as the slow-rising OGLE-2014-SN-131 \citep{Karamehmetoglu2017AA...602A..93K}, the highly luminous ASASSN-14ms ($M_V \approx -20.5$~mag; \citealt{Vallely2018MNRAS.475.2344V, Wang2021ApJ...917...97W}),  the double-peaked iPTF13beo \citep{Gorbikov2014MNRAS.443..671G}, and the long-lasting OGLE-2012-SN-006 \citep{Pastorello2015MNRAS.449.1941P}. The diversity in these observational photometric and spectroscopic properties may indicate a variety of progenitor systems and explosion mechanisms for SNe~Ibn.

The progenitors of SNe~Ibn are usually believed to be massive (17--100 \msun) stars, such as H-poor Wolf-Rayet (WR) stars \citep[e.g.][]{Foley2007ApJ...657L.105F,Pastorello2007Natur.447..829P,Tominaga2008ApJ...687.1208T,Maeda2022ApJ...927...25M}, or for individual SNe~Ibn showing H lines, stars transitioning from luminous blue variable (LBV) to WR stages  \citep[e.g.][]{Smith2012MNRAS.426.1905S, Pastorello2015MNRAS.449.1921P, Reguitti2022A&A...662L..10R}. Although a popular scenario suggests that these are the core-collapse (CC) explosions of very massive stars embedded in helium-rich CSM, there are still many open questions concerning  the homogeneity of Type Ibn progenitors. For example, SNe Ibn have usually been observed in star-forming environments \citep{Taddia2015A&A...580A.131T, Pastorello2015MNRAS.449.1921P}, and thus the massive-star progenitor scenario is favoured. However, this association was challenged by SN Ibn PS1-12sk, which occurred in the outskirts of an elliptical galaxy CGCG 208-042 with no obvious star-formation activity \citep{Sanders2013ApJ...769...39S}.  

In addition to the massive-star scenario, multiple alternative progenitor models are plausible to interpret the observables of Type Ibn events. Based on observations of host-galaxy environments and inspection of explosion sites, lower-mass interacting binaries have also been proposed as progenitor systems \citep[e.g. PS1-12sk, SN~2016jc, and SN 2015G;][]{Sanders2013ApJ...769...39S, Maund2016ApJ...833..128M, Hosseinzadeh2019ApJ...871L...9H, Sun2020MNRAS.491.6000S}; see, in addition, \citet{Samantha2022ApJ...940L..27W} and \citet{Tsuna2024OJAp....7E..82T}.
\citet{Dessart2022A&A...658A.130D} performed numerical simulations to model SN~Ibn spectra, suggesting that a fraction of them can be produced by the explosion of helium-star progenitors exploding in dense CSM.  \citet{Moriya2025PASJ...77.1385M} proposed that some SNe Ibn may stem from ultrastripped SN progenitors that lose substantial mass shortly before their explosion, as a result of violent silicon burning. \citet{Metzger2022ApJ...932...84M} proposed that merger-driven destruction of WR stars rather than a CC explosion can produce SN~Ibn observables. This `explosion'  is actually a disc outflow from the hyperaccretion onto the compact object of the He star. Since the diversity of SNe Ibn is broad, it is perhaps likely that these events originate from multiple formation channels.

In this work, we present a detailed analysis of the photometric and spectroscopic observations of SN\,2024acyl, an SN~Ibn with a linearly declining light curve and early flash-ionisation features. The paper is organised as follows. Section~\ref{sec:information} outlines the discovery, distance, and extinction estimates. The photometric and spectroscopic analysis are presented in Sections~\ref{sec:photometry} and~\ref{sec:spectroscopy}, respectively. Our main results are discussed in Section~\ref{sec:SummaryDiscussion}, and the conclusions are drawn in Section~\ref{sec:Conclusions}.

\section{Discovery, distance, and extinction} 
\label{sec:information}

\begin{figure}
\begin{center}
\includegraphics[width=0.9\linewidth]{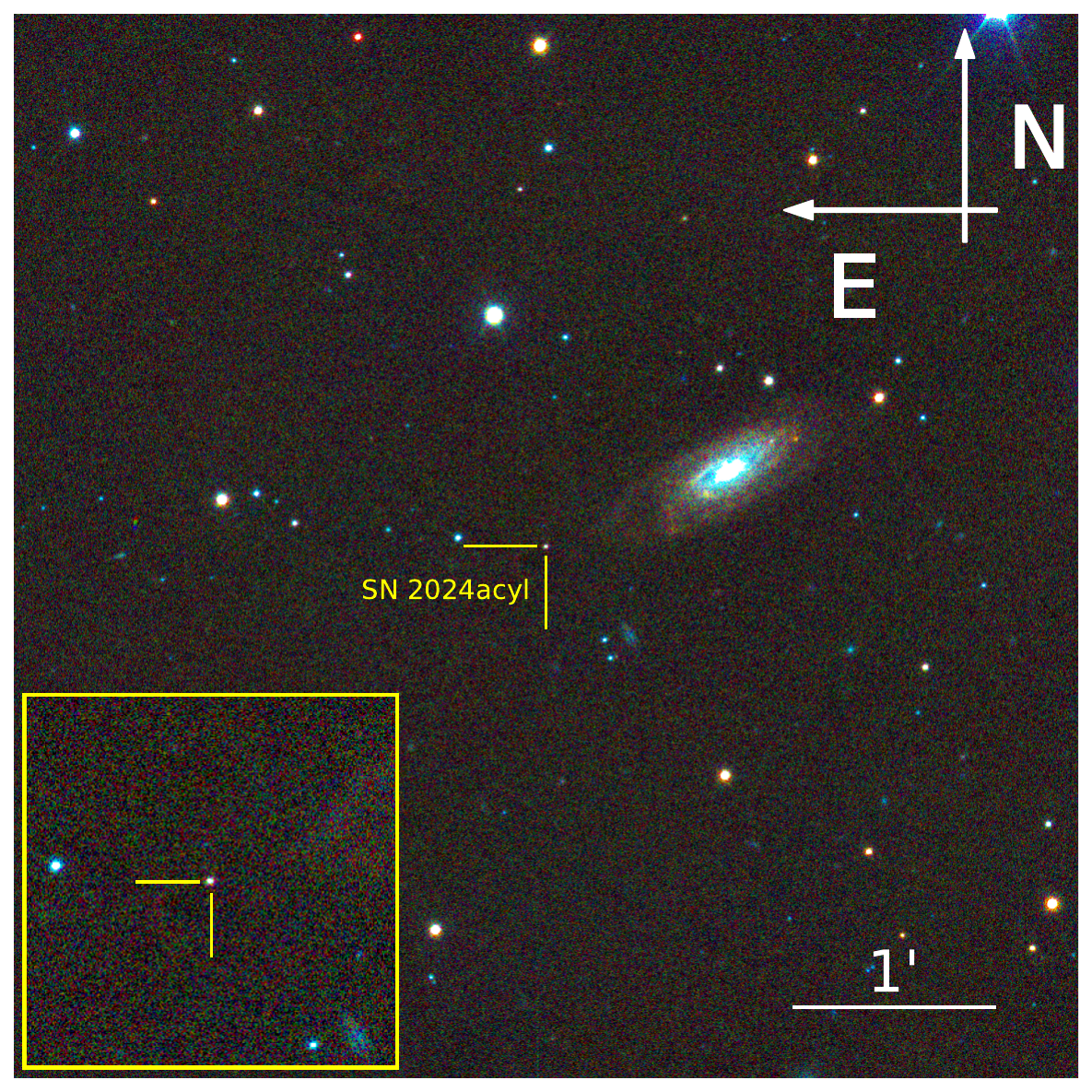}
\caption{SN\,2024acyl in a NOT+ALFOSC coloured image taken with Johnson $B$, $V$, and Sloan $r$ band filters on 2025 January 6. The SN is marked at the crosshair, near the centre of the image.}
\label{fig:finderchart}
\end{center}
\end{figure}

Type Ibn SN\,2024acyl (also known as ATLAS24qxm, GOTO24iwf, and PS24mlb) was first detected by the Asteroid Terrestrial-impact Last Alert System \citep[ATLAS;][]{Tonry2018PASP..130f4505T, Smith2020PASP..132h5002S, Shingles2021TNSAN...7....1S}, on 2024 December 1 UTC (MJD = 60645.28887; UTC dates are used throughout the paper) at the cyan filter magnitude of $c$ = 18.307 mag \citep[AB mag;][]{Tonry2024TNSTR4707....1T}. Early spectra of SN\,2024acyl exhibit prominent He\,\textsc{ii}~$\lambda 4686$ and narrow He~\textsc{i}~$\lambda 5876$ emission lines; hence, it was classified as a young Type Ibn SN with flash features by \citet[][]{Santos2024TNSCR4787....1S} via the extended Public European Southern Observatory (ESO) Spectroscopic Survey of Transient Objects \citep[ePESSTO+;][]{Smartt2015A&A...579A..40S}.

SN\,2024acyl (RA~=~$02^{\rm hr}46^{\rm m}05\fs326$, Dec.~=~$+28\degr01\arcmin17\farcs91$; J2000) has a projected offset of 34 kpc from the core of its probable host galaxy CGCG 505-052, and is 22\farcs98 south and 54\farcs59 east of the galaxy centre (see Fig.~\ref{fig:finderchart}). 
We adopted the host-galaxy redshift from NASA/IPAC Extragalactic Database (NED\footnote{\url{https://ned.ipac.caltech.edu}}) database, $z = 0.026532 \pm 0.000017$ \citep{Springob2005ApJS..160..149S}, which corresponds to a luminosity distance of $d_{L} = 111.2 \pm 7.7$~Mpc and a distance modulus of $\mu_{L} = 35.23 \pm 0.15$~mag. These values are calculated under the assumption of a standard $\Lambda$CDM cosmology with $H_{0} = 73~\mathrm{km\,s^{-1}\,Mpc^{-1}}$, $\Omega_{\mathrm{M}} = 0.27$, and $\Omega_{\Lambda} = 0.73$ \citep[][]{Spergel2007ApJS..170..377S}.
The Milky Way extinction towards SN\,2024acyl is $E(B - V)_{\mathrm{MW}} =$ 0.126 mag  \citep{Schlafly2011ApJ...737..103S},  while the extinction within the host galaxy cannot be firmly constrained owing to the limited spectral resolution and the modest signal-to-noise ratio (S/N) of the SN spectra. Thus, we assume that the total line-of-sight extinction of SN\,2024acyl is equal to the Galactic value, $E(B - V)_{\mathrm{Total}} =$ 0.126 mag, an assumption also supported by the remote location of the SN from the host-galaxy core (see Fig.~\ref{fig:finderchart}). For a detailed analysis of the host environment of SN~2024acyl, we refer to \citet{Dong2025arXiv251103926D}.

\section{Photometry}\label{sec:photometry}

\subsection{Photometric observations}

Soon after the discovery announcement of SN\,2024acyl, we launched a comprehensive multi-band follow-up campaign in the framework of the ePESSTO programme, the Nordic Optical Telescope (NOT) Unbiased Transient Survey 2 (NUTS2\footnote{\url{https://nuts.sn.ie}.}), and other programs. We collected ultraviolet (UV) and optical  photometric data with the facilities listed in Table~\ref{table_telescope} (Appendix~\ref{appendix:facilities}).

\textit{Swift}/UVOT UV and optical data were retrieved from the NASA \textit{Swift} Data Archive\footnote{\url{https://heasarc.gsfc.nasa.gov/cgi-bin/W3Browse/swift.pl}} and measured with the standard UVOT data-reduction pipeline {\tt HEASoft}\footnote{\url{https://heasarc.gsfc.nasa.gov/lheasoft/download.html}} \citep[version 6.19,][]{HEAsoft2014ascl.soft08004N}. The optical photometric data observed from ground-based telescopes were reduced with the dedicated {\sl ecsnoopy}\footnote{{\sl ecsnoopy} is a package for SN photometry using PSF fitting and/or template subtraction developed by E. Cappellaro. A package description can be found at \url{https://sngroup.oapd.inaf.it/ecsnoopy.html}.} pipeline, following  standard procedures as described by \citet[][]{Cai2018MNRAS.480.3424C}. In addition, the 1.6\,m Multi-Channel Photometric Survey Telescope (Mephisto) magnitudes were measured following the methodology presented by \citet{Chen2024ApJ...971L...2C} and \citet{Zou2026ApJ}. We also retrieved archival data from public surveys such as ATLAS and Pan-STARRS \citep[e.g.][]{Chambers2016arXiv161205560C,Flewelling2020ApJS..251....7F,Magnier2020ApJS..251....3M}. ATLAS orange ($o$) and cyan ($c$) magnitudes were processed through the ATLAS Forced Photometry service\footnote{\url{https://fallingstar-data.com/forcedphot/}} \citep{Shingles2021TNSAN...7....1S}, while Pan-STARRS1 (PS1) magnitudes were generated with the PS1 Image Processing Pipeline \citep[IPP;][]{Waters2020ApJS..251....4W, Magnier2020ApJS..251....3M, Magnier2020ApJS..251....6M, Magnier2020ApJS..251....5M}.  The final UV and optical magnitudes of SN\,2024acyl are published at the Strasbourg astronomical Data Centre. 

\subsection{Photometric evolution}
\label{SubSec:PhotometricEvolution}

\begin{figure*}[ht]
    \sidecaption
    \includegraphics[width=12cm]{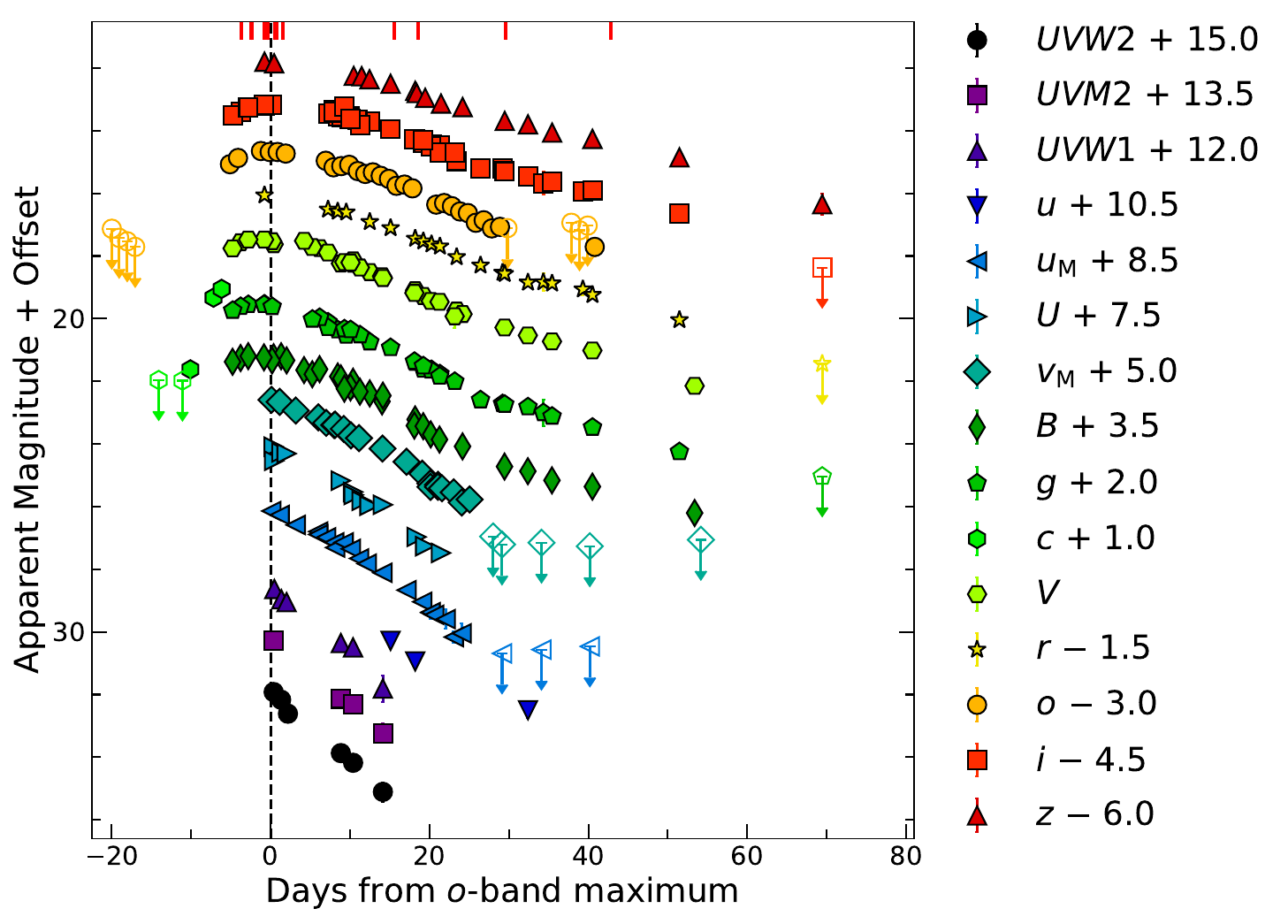}
    \caption{Ultraviolet and optical light curves of SN\,2024acyl. The dashed vertical line indicates the time of the $o$-band maximum light as the reference epoch. The vertical red lines at the top mark spectral observational epochs. The upper limits are plotted with empty symbols with arrows. The light curves for different filters are shifted with arbitrary constants, reported in the legend. The Mephisto $u$- and $v$-band  data points in its unique filter system \citep[for details see][]{Chen2024ApJ...971L...2C, Yang2024ApJ...969..126Y} are indicated by $u_{\mathrm{M}}$ and $v_{\mathrm{M}}$ in the legend.  Magnitude errors are usually smaller than the symbol size.}
    \label{fig:ApparentLCs}
\end{figure*}

The multi-band light curves of SN\,2024acyl are shown in Fig.~\ref{fig:ApparentLCs}. Although its discovery was announced by ATLAS on 2024 December 1 (MJD = 60645.29), an earlier detection on 2024 November 28 (MJD = 60642.39) is recovered in archival images.  While the post-maximum decline is well observed, the pre-maximum evolution is only sparsely covered by the ATLAS $o$- and $c$-band data. Therefore, the explosion time of SN\,2024acyl is estimated from the midpoint between the last non-detection (MJD = 60641.39 in the $c$ band) and the first detection (MJD = 60642.39 in the $c$ band), yielding MJD = 60641.9~$\pm$~0.5 days.
 
To determine the properties of SN\,2024acyl at peak brightness, we performed a polynomial fit on the \textit{o}-band light-curve data around the maximum ($\pm12$ days), which provides a peak magnitude of $17.66 \pm 0.02$ mag and a peak time of MJD = $60652.49 \pm 0.26$. We adopted this epoch as a reference time. The fitting uncertainties were estimated via Monte Carlo simulations. The light curves of SN\,2024acyl are asymmetric, with a relatively long rise ($\sim 10.6$ days) to maximum light, but a fast and linear post-peak decline. We used a linear fit to determine the post-peak decline rates in all bands. The fitted values, which provide a comparison of the fading behaviour across various filters, are reported in Table \ref{decline_rate}. Given the relatively noticeable changes in the slope of the light curves of SN\,2024acyl at approximately +25 days and +45 days, we estimated the decline rates across three distinct time intervals. SN\,2024acyl has a rapid decline compared with other Type Ibn SNe during the whole post-peak evolution in all bands, with the blue-band light curves declining faster than those in the red bands (e.g. two extreme bands of $UVW2$ and $z$ in the first 25 days: $\gamma_{0-25}(UVW2) = 19.8 \pm 1.03$ $\mathrm{mag \, (100\,d)^{-1}}$; $\gamma_{0-25}(z) = 7.98 \pm 0.36$ $\mathrm{mag \, (100\,d)^{-1}}$). 
From 25 d to 45 d, the light curves decline slower for those bands that are visible, with a sort of pseudoplateau (e.g. $\gamma_{25-45}(B) = 5.89 \pm 0.45$ $\mathrm{mag \, (100\,d)^{-1}}$; $\gamma_{25-45}(z) = 5.29 \pm 0.33$ $\mathrm{mag \, (100\,d)^{-1}}$.) Later, the light curve steepens again with a decline rate of $\gamma_{\ge 45} (z) = 8.30 \pm 2.23~\mathrm{mag \, (100\,d)^{-1}}$ until the last detection at about +70 days.

\begin{table}[htbp]
\caption{Decline rates of the multi-band light curves of SN\,2024acyl, along with their uncertainties, in units of mag (100\,d)$^{-1}$.}
\label{decline_rate}
\centering
\begin{tabular}{lccc}
\hline\hline
Filter & $\gamma_{0-25}$ & $\gamma_{25-45}$ & $\gamma_{\ge 45}$\\
\hline
$UVW2$ & $19.80\pm1.03$ & $\cdots$ & $\cdots$\\
$UVM2$ & $20.20\pm4.19$ & $\cdots$ & $\cdots$\\
$UVW1$ & $18.75\pm1.52$ & $\cdots$ & $\cdots$\\
$u$    & $21.19\pm6.94$ & $\cdots$ & $\cdots$\\
$u_{\mathrm{M}}$ & $16.44\pm0.73$ & $\cdots$ & $\cdots$ \\
$U$    & $16.78\pm0.58$ & $\cdots$ & $\cdots$\\
$v_{\mathrm{M}}$ & $13.79\pm0.43$ & $\cdots$ & $\cdots$\\
$B$    & $12.34\pm0.60$ & $5.89\pm0.45$ & $\cdots$\\
$g$    & $11.59\pm0.29$ & $6.44\pm0.47$ & $\ge4.36$\\
$V$    & $11.38\pm0.17$ & $6.79\pm0.37$ & $\cdots$\\
$r$    & $9.63\pm0.29$  & $6.09\pm0.35$ & $\ge7.80$\\
$o$    & $8.97\pm0.57$  & $5.39\pm0.98$ & $\cdots$\\
$i$    & $9.41\pm0.28$  & $5.42\pm0.09$ & $\ge9.64$\\
$z$    & $7.98\pm0.36$  & $5.29\pm0.33$ & $8.30\pm2.23$\\
\hline
\end{tabular}
\end{table}

\begin{figure}
\centering
\includegraphics[width=1\columnwidth]{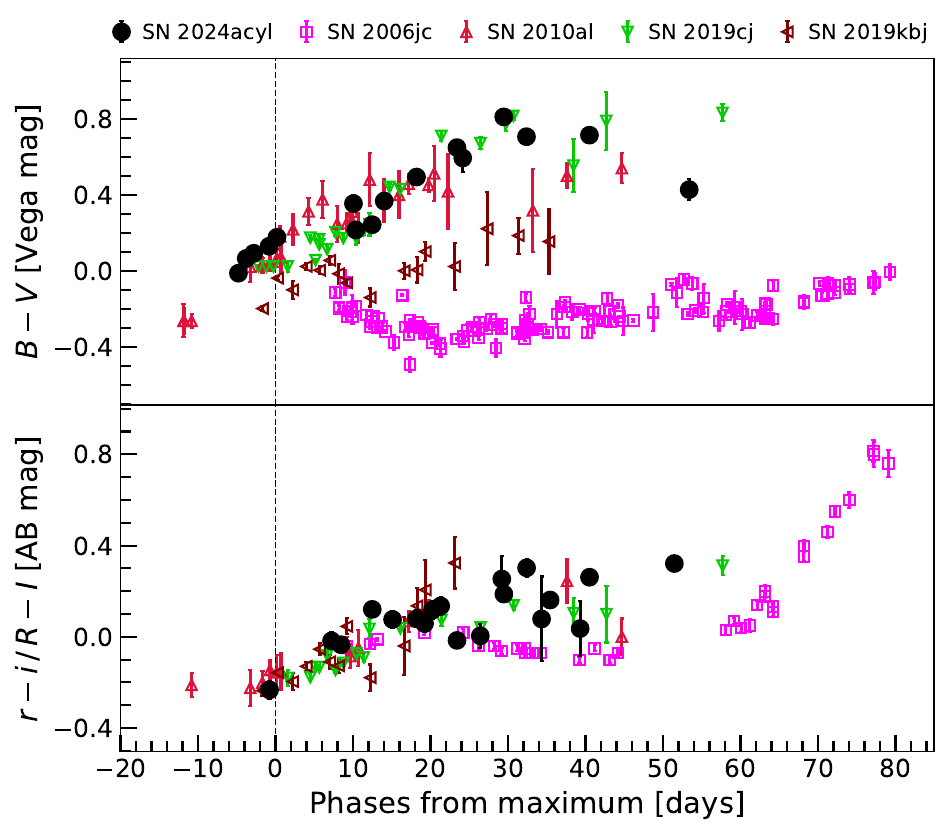}
\caption{Colour evolution of SN\,2024acyl compared with the prototypical Type Ibn SN\,2006jc and other fast, linearly declining SNe Ibn. {\it Top panel:} $B~-~V$ colours. {\it Bottom panel:} $R~-~I$ or $r~-~i$ colours. The colour curves have been corrected for galactic extinction.}
\label{fig:colour}
\end{figure}

The colour evolution of SN\,2024acyl is shown in Fig. \ref{fig:colour}, compared with those of a selected sample of Type Ibn SNe, including the prototypical Type Ibn SN\,2006jc and a few objects\footnote{The comparison SNe Ibn include SNe\,2006jc \citep{Pastorello2007Natur.447..829P}, 2010al \citep{Pastorello2015MNRAS.449.1921P}, 2019kbj \citep{Ben-Ami2023ApJ...946...30B}, and 2019cj \citep{Wang2024A&A...691A.156W}.} that share similar light-curve properties with SN\,2024acyl. In the top panel of Fig. \ref{fig:colour}, the $B~-~V$ colour becomes red very rapidly, moving from a blue colour of $+$0.0 mag at $-$5 days to a red one of $+$0.8 mag at $+$30 days. After maximum brightness, the $B~-~V$ colour turns bluer to $+$0.4 mag in the following days up to $+$55 days. However, the comparison objects reveal that there is diversity in the colour evolution of SNe Ibn. The $B~-~V$ colour of SNe\,2010al and 2019cj become redder rapidly at their early stages resembling that of SN\,2024acyl, but the later evolution turns to moderately redder colours. SN\,2019kbj shows a similar trend but its colour is bluer.  SN\,2006jc evolves to a blue colour from $-$0.1 mag to $-$0.4 mag in its first $+$15 days and gradually becomes redder at around $-$0.2 mag (with minor fluctuations) until day $+$60. Subsequently, it rapidly became redder, which is likely to associated with dust formation \citep{Mattila2008MNRAS.389..141M, Smith2008ApJ...680..568S, DiCarlo2008ApJ...684..471D}. In the bottom panel of Fig. \ref{fig:colour}, the $r~-~i$ colour of SN\,2024acyl slowly increases from $-$0.2 mag at $\sim$0 day to $+$0.1 mag at $+$15 days and subsequently settles to about $+$0.3 mag ($+$51 days) but with some fluctuations. We caution that there is also the possibility that fluctuations in the colour curves are likely due to data quality and are not of the SN.
The evolution of the $R~-~I$ / $r~-~i$ colour in the comparison objects is consistent with the trend seen in SN\,2024acyl within the observed time window, although the colour of SN\,2006jc becomes much redder at late phases.

\begin{figure}[htp]
\begin{center}
\includegraphics[width=1\columnwidth]{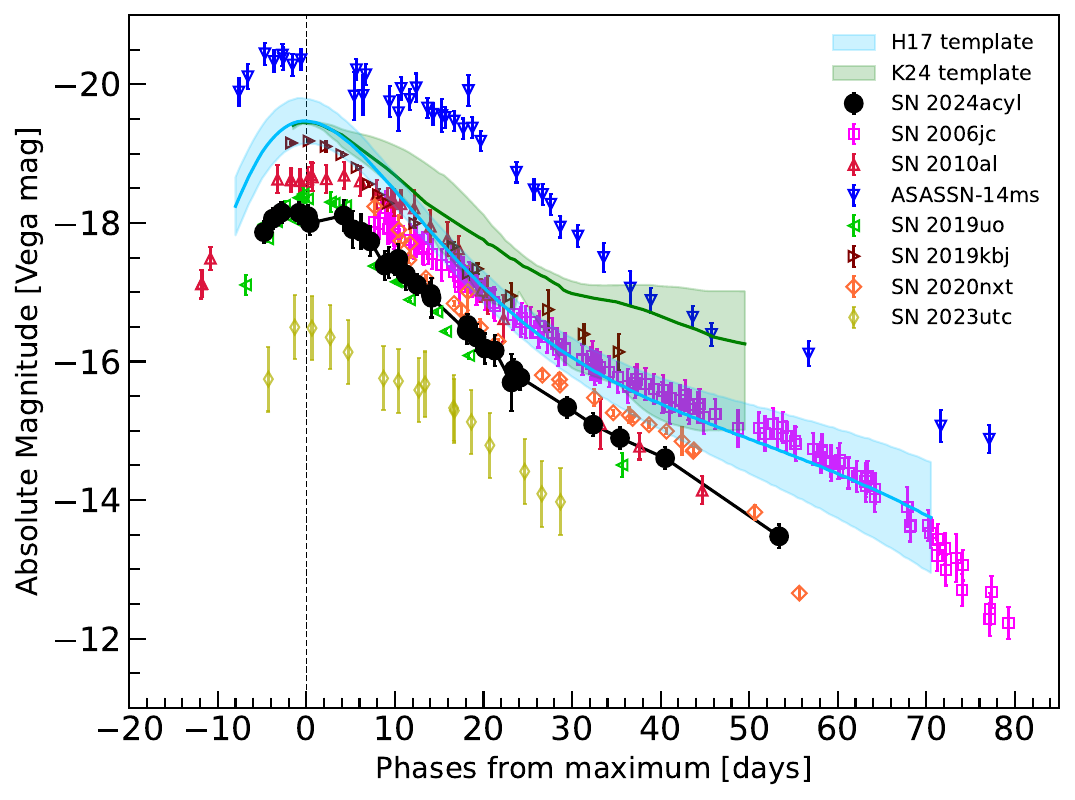}
\caption{Light curves in the $V$-band of SN\,2024acyl, including the comparison SNe Ibn. Template $V$-band light curves for Type~Ibn SNe are from \citet[][blue]{Hosseinzadeh2017ApJ...836..158H} and \citet[][green]{Khakpash2024ApJS..275...37K}. Due to data-coverage limitations, SN~2023utc is represented using $r$-band photometry converted to the Vega system.}
\label{fig:AbsoluteMag}
\end{center}
\end{figure}

Adopting the distance and reddening estimates reported in Sec. \ref{sec:information}, SN\,2024acyl reached absolute magnitudes of $M_{B} = -18.02 \pm 0.15 \, \mathrm{mag}$, $M_{g} = -18.13 \pm 0.15 \, \mathrm{mag}$, $M_{V} = -18.14 \pm 0.15 \, \mathrm{mag}$, $M_{o} = -17.88 \pm 0.15 \, \mathrm{mag}$, and $M_{i} = -17.82 \pm 0.15 \, \mathrm{mag}$. Only upper limits can be estimated for other bands owing to the incomplete data coverage around maximum: $M_{r} < -18.04$ mag and $M_{z} < -17.59$ mag. 
SN\,2024acyl is slightly fainter than the average absolute magnitude of SNe Ibn \citep[$M_r \approx -19~\mathrm{mag}$;][]{Pastorello2016MNRAS.456..853P,Hosseinzadeh2017ApJ...836..158H,Wang2025A&A...700A.156W}, and much fainter than the highly luminous ASASSN-14ms \citep[$M_V \approx -20.5$~mag;][]{Vallely2018MNRAS.475.2344V, Wang2021ApJ...917...97W}. On the other hand, SN\,2024acyl is much brighter than SN~2023utc, which is the faintest Type~Ibn SN reported in the literature \citep[$M_{r} = -16.4 \pm 0.5$ mag; ][]{Wang2025A&A...700A.156W}. To highlight the fast and linear post-peak photometric decline of SN\,2024acyl, we compared the $r$-band light curve of SN\,2024acyl with those of a few representative SNe Ibn and the Type Ibn templates presented by \citet{Hosseinzadeh2017ApJ...836..158H} and \citet{Khakpash2024ApJS..275...37K} (see Fig. \ref{fig:AbsoluteMag}). The light curve of SN\,2024acyl declines rapidly at phases later than $+$5 days, consistent with most SNe Ibn (including SNe\,2006jc and 2020nxt), and it follows the behaviour of the templates released by \citet[][]{Hosseinzadeh2017ApJ...836..158H}.  However, SN\,2024acyl before $+$5 days is less luminous than other SNe Ibn whose peak absolute magnitudes range from $-18.9$~mag to $-20.5$~mag, with the notable exception of SN~2023utc.

To make a meaningful comparison of SN\,2024acyl with other SNe Ibn, we constructed their pseudobolometric light curves based on the photometry available in the same set of filters,  from the $B$ to the $I/i$ bands. Therefore, we first converted extinction-corrected magnitudes to flux densities, and then integrated the spectral energy distribution (SED) within their effective wavelengths. In our computation, we made the assumption that the flux contribution outside the integration limits, which represent the coverage of each bands, is zero. Occasionally, when some photometric data in a given filter were not available, we interpolated or extrapolated the missing flux from the nearest available photometry assuming a constant colour evolution. 

\begin{figure}[htp]
\includegraphics[width=1\columnwidth]{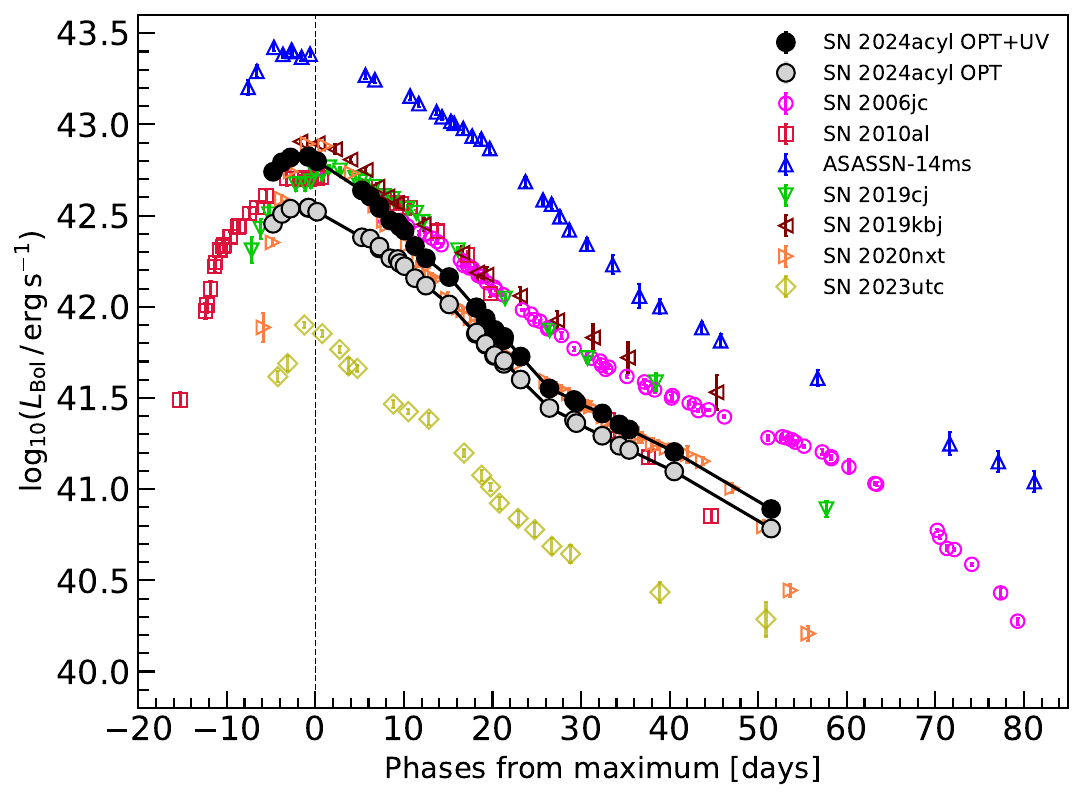}
\caption{Pseudobolometric light curves of SN\,2024acyl and the comparison SNe Ibn. The comparison objects have luminosities comparable to that of the `optical' luminosity of SN\,2024acyl, integrating from the $B$ to the $I/i$ bands.}
\label{fig:quasibolom}
\end{figure}

The resulting pseudobolometric light curves of SN\,2024acyl and the compared SNe Ibn are shown in Fig.~\ref{fig:quasibolom}. 
The pseudobolometric light curve of SN\,2024acyl is broadly similar to other Type Ibn events. The peak  `optical' luminosity of SN\,2024acyl, ($3.5\pm0.8) \times 10^{42}$ erg s$^{-1}$, is between those of most SNe Ibn ($\sim$3 to 20 $\times 10^{42}$ erg s$^{-1}$) and the faintest SN 2023utc ($\sim 7.1 \times 10^{41}$ erg s$^{-1}$). The peak  `UV+Optical' luminosity, ($6.7\pm0.4) \times 10^{42}$ erg s$^{-1}$, of SN\,2024acyl is still fainter than that of ASASSN-14ms ($\sim 2.3 \times 10^{43}$ erg s$^{-1}$). We observed a significant difference in peak luminosity between the `optical' and `UV+Optical' results. This indicates that the contribution from UV bands is significant during the early phases, which is consistent with observations of other SNe Ibn \citep[see][]{Wang2024MNRAS.530.3906W, Wang2024A&A...691A.156W}. Such a feature suggests a high-temperature scenario in the early phases where the peak of the SED falls within the UV bands (e.g. when $T_{\mathrm{BB}}=20,000$~K, $\lambda_{\mathrm{max}}\approx1500$~\AA). Furthermore, strong interaction may generate energetic photons; as the ejecta are opaque to this radiation in the early phases, these photons are thermalised into softer UV emission. These factors combined explain the high contribution of UV bands in the early phases, consistent with the work of \cite{Maeda2022ApJ...927...25M}. In addition, we estimated the radiated energies of SN\,2024acyl from the `optical' and `UV+Optical' pseudobolometric light curves, using a non-parametric fit of a~\texttt{ReFANN}\footnote{\url{https://github.com/Guo-Jian-Wang/refann}} code \citep[see details in][]{Wang2020ApJS..249...25W, Wang2020ApJS..246...13W, Wang2021MNRAS.501.5714W}. The resulting radiated energies integrated with the entire photometric evolution time are $(5.0\pm0.4) \times 10^{48}$ erg and $(8.5\pm0.6) \times 10^{48}$ erg, respectively. These values of SN\,2024acyl are within the range of $(1$–$32) \times 10^{48}$ erg, as reported for the typical Type Ibn sample \citep[see Table 2 of][]{Wang2025A&A...700A.156W}. Note that these values should be considered lower limits owing to limited temporal and wavelength coverage. 

\subsection{multi-band light-curve modelling} 
\label{Sec:MOSFiT}

\begin{figure*}[htbp]
\sidecaption
\includegraphics[width=12cm]{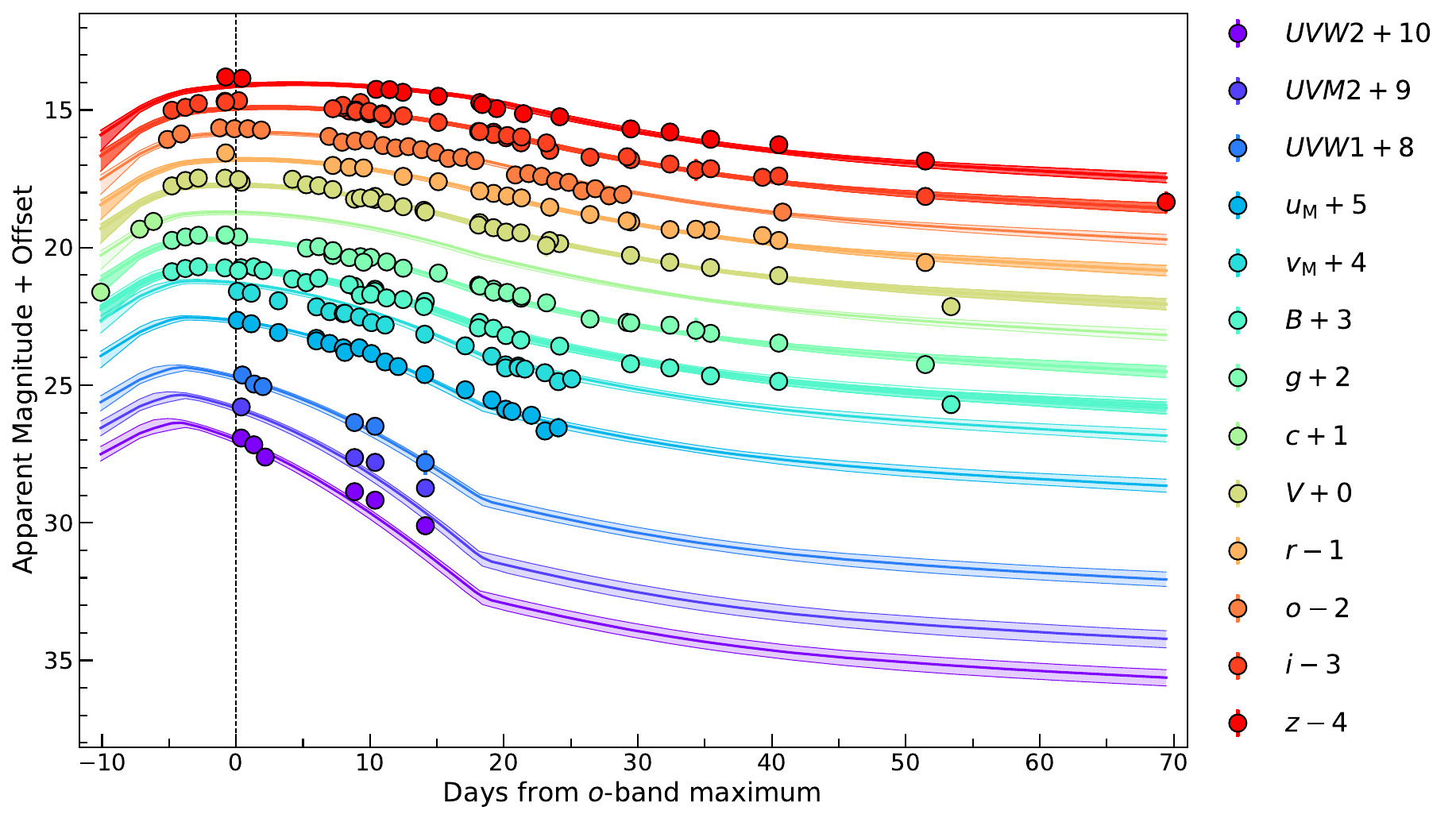}
\caption{Fits to the multi-band light curves of SN~2024acyl with $n=10$, $\delta=0$, and $s=0$ using the \texttt{MOSFiT} Monte Carlo code with the hybrid \texttt{Ni+CSM} model. For each filter, a subset of randomly sampled model light curves from the posterior distributions are displayed to illustrate the uncertainty of the model fits.}
\label{fig:mosfit_lc}
\end{figure*}

Type Ibn SNe are characterised by strong interaction between their ejecta and a helium-rich CSM \citep{Karamehmetoglu2017AA...602A..93K, Kool2021A&A...652A.136K, Pellegrino2022ApJ...926..125P}. This ejecta-CSM interaction (CSI) is a dominant power source for the light curve, necessitating a more complex model than one based solely on  radioactive decay (RD). Therefore, to accurately model the light curve of SN 2024acyl, we adopted a hybrid model that combines contributions from both RD and CSI, following  \citet{Chatzopoulos2012ApJ...746..121C}.

We adopted the \texttt{MOSFiT} Monte-Carlo fitting code \citep{Guillochon2018ApJS..236....6G}, widely used in other Type Ibn SNe \citep[e.g.][]{Kool2021A&A...652A.136K, Farias2024ApJ...977..152F}, to fit the multi-band light curve of SN~2024acyl. The bands we used to constrain the explosion parameters here were the UV plus the optical $u_{\mathrm{M}}v_{\mathrm{M}}$\textit{BgcVroiz} ones. The data in the $u_{\mathrm{M}}$ and $v_{\mathrm{M}}$ bands (for which the effective wavelengths are 3454\,\AA\ and 3854\,\AA, respectively) are taken from the Mephisto survey with its unique filter system, which has  better coverage around the peak and in the post-maximum phases. Using \texttt{MOSFiT}, we obtained the posterior distribution of each parameter along with its uncertainty, and inspected the degeneracy between different parameters.  
Furthermore, by fitting the light curve of each band independently using \texttt{MOSFiT}, we can accurately model the colour evolution of SN~2024acyl. This approach is particularly effective for the refined $u_\mathrm{M}$- and $v_\mathrm{M}$- bands from the Mephisto survey.

The RD-CSI model implemented in the \texttt{MOSFiT} code is based on the formalism of \cite{Chatzopoulos2012ApJ...746..121C}. In this semi-analytical framework, the luminosity, $L(t)$, is computed as the diffusion of an energy input through the ejecta:
\begin{equation}
    \begin{split}
    L(t) = \left[\frac{1}{t_0} e^{-\frac{t}{t_0}}\int_{0}^{t}\mathrm{d}\tau\, e^{\frac{\tau}{t_0}}L_{\mathrm{inp}}(\tau)+\frac{E_\mathrm{init}}{t_0}e^{-\frac{t}{t_0}}\right] \\
    \times \left(1-e^{-\kappa_\gamma \rho(t)R(t)} \right) \, . 
    \end{split}
\end{equation}
Here, $t_{0} = \kappa M_{\rm{ej}}/\beta c v_{\rm{ph}}$ is the characteristic diffusion timescale, where $\beta \approx 13.8$ is an integration constant derived from the diffusion model \citep[see e.g.][]{Arnett1982ApJ...253..785A,Valenti2008MNRAS.383.1485V}. The input energy source, $L_{\rm{inp}}(\tau)$, is the sum of RD and CSI. The RD component is powered by the RD chain $^{56}$Ni$\rightarrow^{56}$Co$\rightarrow^{56}$Fe. The CSI component consists of contributions from both a forwards and a reverse shock, the strengths of which depend on the density profiles of the ejecta and the surrounding CSM \citep{Chevalier1982ApJ...258..790C}.
The model is described by several key parameters. The diffusion process is primarily governed by the $^{56}$Ni fraction ($f_{\rm{^{56}Ni}}$), the ejecta mass ($M_{\rm{ej}}$), the ejecta kinetic energy ($E_{\rm{k}}$), the optical opacity ($\kappa$), and the gamma-ray opacity ($\kappa_{\gamma}$, which accounts for gamma-ray leakage). The CSI is characterised by the CSM mass ($M_{\rm{CSM}}$), its inner radius ($R_0$), and parameters describing the density profiles of the CSM ($s$) and the ejecta ($n$, $\delta$). The CSM density profile is expressed as $\rho_{\rm{CSM}} \propto r^{-s}$, where $s=2$ corresponds to a steady stellar wind, while $s=0$ represents a dense, shell-like CSM with constant density. Additionally, the \texttt{MOSFiT} fitting process includes some additional parameters including the explosion time ($t_{\rm{exp}}$), a minimum photospheric temperature ($T_{\rm{min}}$), and an uncertainty term ($\sigma$) added in quadrature to the observational errors to account for model uncertainties, which indicate the fitting quality. 

The complexity of the RD-CSI model and the high dimensionality of its parameter space make the fitting process computationally challenging. To simplify the problem, we fixed several parameters or use more restrictive priors based on physically motivated assumptions. 
Given that the fitting results are insensitive to the density profile of the ejecta \citep[see][]{Villar2017ApJ...849...70V, Kool2021A&A...652A.136K, Farias2024ApJ...977..152F}, we use restrictive priors for these parameters. For the density profile of the inner ejecta, we adopt a constant-density profile ($\delta=0$; \citealt{Wang2025A&A...700A.156W}). Regarding the density profile of the stellar envelope, \cite{Chatzopoulos2012ApJ...746..121C} noted that for the relatively compact scenario, the index $n$ needs to be smaller; specifically, $n\lesssim10$ for BSG and WR cases, which represent a radiative envelope, in contrast with the convective envelope of RSG stars ($n\approx12$). Furthermore, \cite{Farias2025ArXiv} noted that for the RD+CSI hybrid model, the best-fit $n$ is approximately 9. Conversely, \cite{Kool2021A&A...652A.136K} and \cite{Wang2025A&A...700A.156W} fixed $n=12$ as a standard value, as it does not significantly affect the result. Thus, considering all these scenarios, we adopted $n=10$ and $n=12$, which represent compact and generally used cases, respectively. For the density profile of the CSM, we considered two scenarios: $s=0$ and $s=2$. The former represents a shell-like CSM driven by strong WR winds or eruptive mass-loss events, while the latter represents a steady stellar wind \citep{Chatzopoulos2012ApJ...746..121C}.
We also adopted a constant optical opacity of $\kappa = 0.1 \, \mathrm{cm^2 \, g^{-1}}$, typical for helium-rich ejecta \citep{Prentice2019MNRAS.485.1559P}, and a gamma-ray opacity of $\kappa_\gamma = 0.027 \, \mathrm{cm^2 \, g^{-1}}$. This procedure reduced the number of free parameters in our fit to eight: $f_{^{56}\mathrm{Ni}}$, $M_{\mathrm{ej}}$, $E_{\mathrm{k}}$, $M_{\mathrm{CSM}}$, $\rho_\mathrm{CSM}$, $R_0$, $T_{\rm{min}}$, and the additional uncertainty term $\sigma$. Given the potential for a complex posterior distribution and the high-dimension parameter space, we employed the nested sampling algorithm implemented in the \texttt{dynesty} package \citep{Speagle2020MNRAS.493.3132S} instead of traditional ensemble-based samplers. We initialised the sampler with 120 live points (`walkers') and ran the algorithm iteratively until the stopping criterion was reached to ensure the good convergence of the samplers.

The best-fit model is presented in Fig.~\ref{fig:mosfit_lc}, overlaid on the multi-band photometric data\footnote{Since the light curve is not sensitive to the density-profile parameters ($n$, $\delta$), and the parameter $s$ has only a minor effect on the light-curve shape, in Fig.~\ref{fig:mosfit_lc} we present only a single best-fit scenario with $n=10$, $\delta=0$, and $s=0$.}. The corresponding model parameters are listed in Table~\ref{apptab:lcmodeling} (Appendix \ref{appendix:LCmodel_parameters}). All parameters were constrained by the observational data, with uncertainties defined by the 68\% ($\sim1\sigma$) confidence intervals of their posterior distributions. The additional uncertainty added to the observation data is $\sim0.15 \, \mathrm{mag}$ in this fitting, consistent with the large-sample analysis of \cite{Wang2025A&A...700A.156W}. 

Therefore, the results should be regarded as indicative and used with caution. The minimum photospheric temperature, $T_{\mathrm{ph, \, min}}$, is not listed, as the model is insensitive to this parameter compared to other key parameters \citep{Nicholl2017ApJ...850...55N}. The corner plot, illustrating the posterior distributions and the degeneracies between parameters, is shown in Fig.~\ref{fig:modfit_corner1} in Appendix~\ref{appendix:corner}. 

We found that for both $n=12$ and $n=10$, the models with $s=0$ provide tighter constraints on the parameters compared to the $s=2$ models. Furthermore, the Bayesian evidence of the $s=0$ models is approximately $\mathrm{log}\mathcal{Z}\approx304$, whereas the $s=2$ models yield $\mathrm{log}\mathcal{Z}\approx296$. This indicates that the $s=0$ models are statistically favoured. The value of $n$ does not significantly affect the key parameters; both $n=12$ and $n=10$ yield similar low-mass solutions (with consistent $M_{\mathrm{CSM}}$ and density) and share comparable Bayesian evidence ($\Delta\mathrm{log}\mathcal{Z}\lesssim1$), which is consistent with \cite{Kool2021A&A...652A.136K} and \cite{Wang2025A&A...700A.156W}. Considering these findings, we indicate that the posterior distributions of the parameters are well constrained.

As presented in Fig.~\ref{fig:modfit_corner1}, $M_{\mathrm{ej}}$ and $E_{\mathrm{k}}$ are degenerate. Thus, we can robustly constrain the $M_{\mathrm{ej}} \lesssim 1~\msun$ and $E_{\mathrm{k}} \lesssim 0.1\times10^{51}~\mathrm{erg}$. The choice of the density profile index $n$ introduces a systematic uncertainty of $\sim0.2\,\msun$ for  $M_{\mathrm{ej}}$ and $\sim0.02\times10^{51}\,\mathrm{erg}$ for  $E_{\mathrm{k}}$. By considering that most SNe Ibn exploded from relatively compact progenitors, we take the posterior of $n=10$ scenario as a reasonable result. The best-fit values are $M_{\mathrm{ej}} = 0.49^{+0.11}_{-0.09} \, \msun$ and $E_{\mathrm{k}} = 0.06^{+0.01}_{-0.01} \times 10^{51}\,\mathrm{erg}$ for SN~2024acyl. 
The value of $M_{\mathrm{ej}}$ is comparable to those of other Type Ibn SNe such as PS1-12sk ($\sim$0.3\,\msun; \citealt{Sanders2013ApJ...769...39S}), located at the lower end of the $M_{\mathrm{ej}}$ distribution of SNe Ibn. 
The value of $E_{\mathrm{k}}$ is reasonable because it falls in the low-energy tail of SNe Ibn, which is $(0.06 - 0.91) \times 10^{51} \, \mathrm{erg}$. However, we caution that these are correlated, as a higher ejecta mass results in a higher kinetic energy, and vice versa.

The derived properties of the CSM are also typical of SNe Ibn. The CSM properties are not significantly affected by the choice of $n$, which implies that they are robust against variations in the density profile. The $M_{\mathrm{CSM}}$ of SN~2024acyl is $0.51^{+0.05}_{-0.04} \, \msun$, which is comparable to that of iPTF15ul and SN~2019uo with the assumption of shell-like CSM ($s=0$) ($\sim0.3$--0.7\,$\msun$; \citealt{Pellegrino2022ApJ...926..125P, Gangopadhyay2020ApJ...889..170G}). The best-fit inner radius of the CSM of SN~2024acyl is $17.8^{+3.6}_{-3.0}\,\mathrm{AU}$, which falls within the observed range of $9-60$ AU, bracketed by examples such as SN~2020nxt ($\sim9$ AU; \citealt{Wang2025A&A...700A.156W}) and iPTF15ul ($\sim60$ AU; \citealt{Pellegrino2022ApJ...926..125P}). The posterior CSM density for SN~2024acyl is $\rho_{\mathrm{CSM}} = (8.3_{-1.2}^{+2.7})\times10^{-12} \, \mathrm{g\,cm^{-3}}$. The CSM properties of SN~2024acyl is consistent with that of other SNe Ibn with $s=0$.
The outer radius of the CSM can be derived from $R_{0}$, $M_{\mathrm{CSM}}$, and $\rho_{\mathrm{CSM}}$, which is around 24.3~AU, indicating a relatively thin CSM with a thickness of $\sim$6.5~AU, suggesting eruptive mass loss of the progenitor. 
Therefore, the reasonable posterior distributions of the CSM parameters (e.g. $R_0$, $M_{\mathrm{CSM}}$, and $\rho_{\mathrm{CSM}}$), and the higher Bayesian evidence $\mathrm{log}\mathcal{Z}\approx304$ for $s=0$ models, indicate that its properties are well constrained. The profile of the CSM can be a probe of the mass-loss history of the progenitor, as discussed in Sec.~\ref{sec:physics_2024acyl}.

For interacting SNe, the contribution from RD is often secondary, with  $M_{\mathrm{Ni}}$ typically being low ($\lesssim0.1\,\msun$; \citealt{Maeda2022ApJ...927...25M, Ben-Ami2023ApJ...946...30B}). For SN~2024acyl, considering the posteriors of mass fraction of $^{56}$Ni and the ejecta mass, we find $M_{\mathrm{Ni}}=0.018\,\msun$. This value is comparable to that of SN~2019wep ($0.015\pm0.05 \, \msun$; \citealt{Pellegrino2022ApJ...926..125P}) and lies within the range (0.001--0.15\,$\msun$) of the sample from \cite{Wang2025A&A...700A.156W}. Finally, we estimated a characteristic ejecta velocity using the relation $v_{\mathrm{ej}}\approx\sqrt{2 E_{\mathrm{k}}/M_{\mathrm{ej}}}$. This yields a velocity of $\sim3500 \, \mathrm{km \, s^{-1}}$ for SN~2024acyl, comparable to that of SN~2020bqj ($\sim3300\,\mathrm{km\,s^{-1}}$; \citealt{Kool2021A&A...652A.136K}) and SN~2020taz ($\sim3480\,\mathrm{km\,s^{-1}}$; \citealt{Wang2025A&A...700A.156W}). However, as a characteristic value, the velocity derived here may differ from the actual spectroscopic velocity and should therefore be used with caution.

\section{Spectroscopy}\label{sec:spectroscopy}

\subsection{Spectroscopic observations}
\label{subsec:SpectroData}

Our spectroscopic observations of SN\,2024acyl were obtained using multiple instrumental configurations: The 2.4\,m LJT equipped with the Yunnan Faint Object Spectrograph and Camera \citep[YFOSC;][]{Wang2019RAA....19..149W} at Gaomeigu, Lijiang, China; the Kast double spectrograph \citep{Miller1993} mounted on the 3\,m Shane telescope at the Lick Observatory; the 3.58\,m New Technology Telescope (NTT) equipped with the ESO Faint Object Spectrograph and Camera 2 \citep[EFOSC2;][]{Buzzoni1984Msngr..38....9B}, located at La Silla, Chile; the 3.58\,m Telescopio Nazionale Galileo (TNG) with the Device Optimized for the LOw RESolution \citep[DOLORES;][]{Molinari1997MmSAI..68..231M} spectrograph, hosted on La Palma, Spain; and the 8\,m Gemini–North telescope equipped with Gemini Multi‐Object Spectrograph (GMOS‐N; \citealt{Hook2004PASP..116..425H, Gimeno2016SPIE.9908E..2SG}) on Mauna Kea in Hawai'i, USA. Additionally, we collected a single-epoch (2024-12-04) spectrum from the Transient Name Server (TNS\footnote{\url{https://www.wis-tns.org/object/2024acyl}}), which was obtained by \citet{Soubrouillard2024TNSCR4856....1S}. Basic information for the spectra is reported in Table~\ref{table:speclog_2024acyl} (Appendix~\ref{Spec_log}). 

Reduction of the GMOS-N data was done by \citep[\texttt{DRAGONS}][]{Labrie2023RNAAS...7..214L} packages following standard procedures. 
The Shane/Kast spectrum was taken at or near the parallactic angle to minimise slit losses caused by atmospheric dispersion \citep{Filippenko1982PASP...94..715F}. Its data reduction followed standard techniques for CCD frame processing and spectrum extraction \citep{Silverman2012MNRAS.425.1789S} using {\sc iraf} routines and custom \textsc{Python} and IDL codes\footnote{\url{https://github.com/ishivvers/TheKastShiv}}. 
The NTT/EFOSC2 spectra were reduced using a dedicated pipeline \texttt{PESSTO}\footnote{\url{https://github.com/svalenti/pessto}} \citep{Smartt2015A&A...579A..40S}, while spectra obtained from other instruments were processed using standard procedures in the \texttt{IRAF} environment. Specifically, the raw data were first pre-reduced with preliminary steps, such as bias, overscan, trimming, and flat-fielding corrections. Then, we extracted one-dimensional spectra from the two-dimensional frames. Wavelength and flux calibrations were performed using spectra of comparison lamps and spectrophotometric standard stars, respectively, which were observed during the same night and with the same instrumental configurations as the SN spectra. The wavelength-calibrated spectra were cross-checked with the night-sky emission lines, while the flux-calibrated spectra were improved by calibrating to the coeval broadband photometry. We also examined the consistency between the colours derived from synthetic photometry and the observed colour evolution. However, we noted a minor discrepancy in the data, as the difference in the $g-r$ and $r-i$ bands is approximately $\sim0.1$~mag. Since we could not determine the precise reddening parameters, we treated this discrepancy as a systematic uncertainty of the extinction in our fitting procedure. Finally, the strongest telluric absorption bands (e.g. $\mathrm{O}_2$ and $\mathrm{H}_2\mathrm{O}$) in the SN spectra were removed using the spectra of standard stars. 

\subsection{Spectroscopic evolution}
\label{subsec:SpectroEvolution}

\begin{figure*}
\begin{center}
\includegraphics[width=1.8\columnwidth]{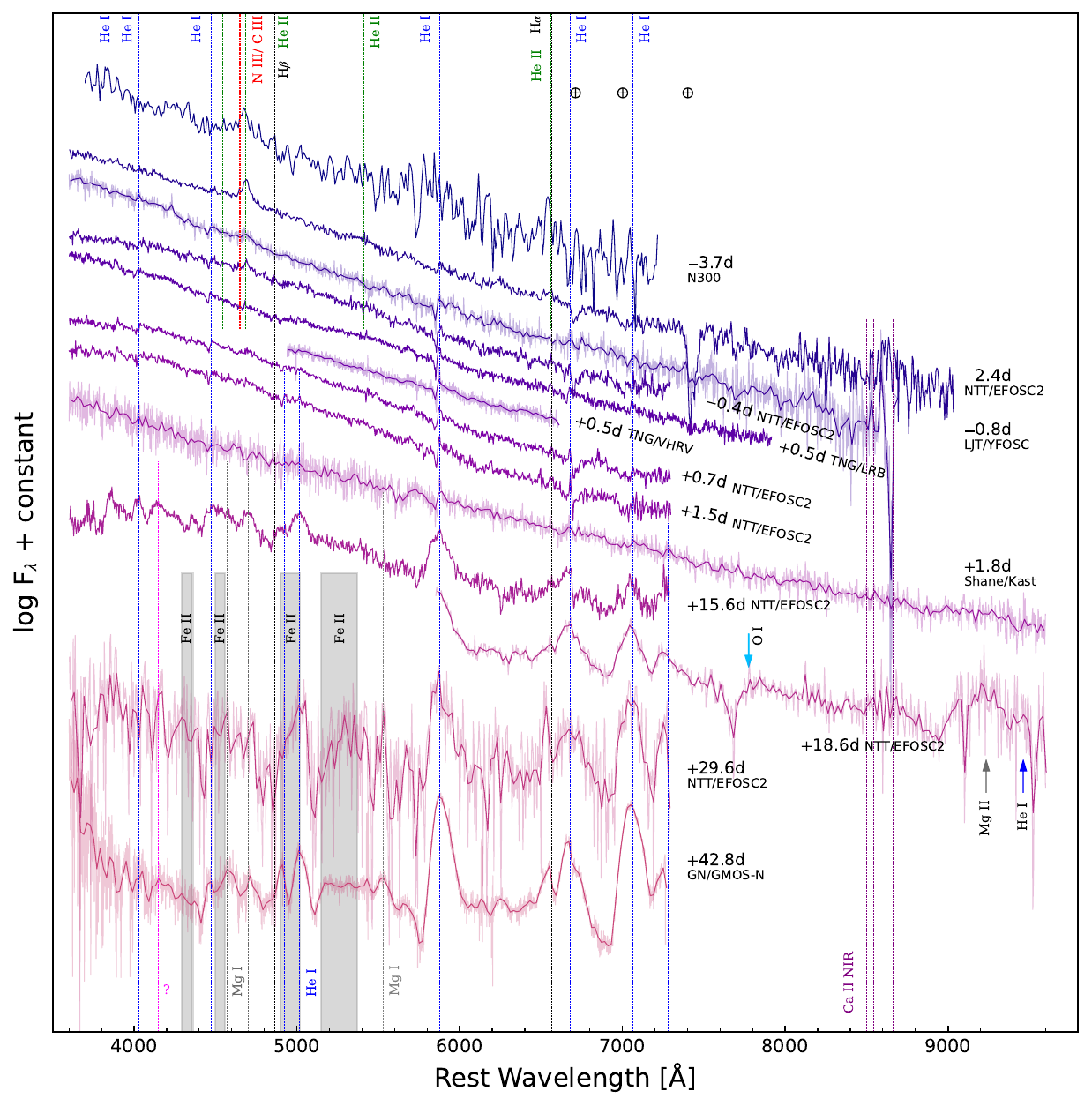}
\caption{Time sequence of SN\,2024acyl spectra. Some prominent features, such as \Hei, \Heii, and \Niii,  are marked with vertical lines, while the strongest telluric absorption bands are indicated with the $\oplus$ symbols. The phases reported to the right of each spectrum are from the epoch of $o$-band maximum light (MJD = $60652.49 \pm 0.26$; 2024-12-08). Spectra with a low S/N were binned with 20~\AA\ each bin; the original (unbinned) spectra are displayed in lighter colours behind. Reddening and redshift corrections have been applied to the spectra.}
\label{fig:SN_spectra}
\end{center}
\end{figure*}

We collected 12 optical spectra of SN\,2024acyl, spanning about 50 days and covering all crucial phases of its evolution. The spectral evolution of SN\,2024acyl, as shown in Fig.~\ref{fig:SN_spectra}, exhibits features typical of SNe Ibn. 

The first spectrum of SN\,2024acyl was obtained on 2024 December 4 (phase $\sim -$3.7 days from  maximum light; \citealt{Soubrouillard2024TNSCR4856....1S}). This low-S/N spectrum shows a blue, almost featureless continuum and did not provide a secure classification. A prominent bump detected at $\sim 4600$--4700\,\AA\ is probably due to a blend of \Ciii~$\lambda$4648, \Niii~$\lambda$4640, and \Heii~$\lambda$4686 emission lines. We estimated the temperature  by fitting the continuum with a blackbody function. Since we noticed that the systematic error is likely to be underestimated, we introduce the jitter term as the additional error in the fitting process. The best-fit results yield a photospheric temperature of $T_{\rm BB} = 19,200_{-1,100}^{+1,200}~{\rm K}$ with the additional jitter $\mathrm{log}(\sigma/10^{42}~\mathrm{erg\,s^{-1}})=-4.45_{-0.07}^{+0.08}$.\footnote{Since we added the additional jitter term $\mathrm{log}(\sigma)$, it explicitly accounts for the underestimated systematic errors and intrinsic deviations from the model statistically.} Subsequently, the second spectrum of SN\,2024acyl ($\sim$ $-$2.4 days) supports the classification of this event as a Type Ibn SN (see Sec.~\ref{sec:information}). This spectrum is still dominated by a blue continuum ($T_{\rm BB} = 15,000\pm600~{\rm K}$), but now the \Hei~$\lambda$5876 line is clearly detected with a narrow P-Cygni profile. The position of the minimum of this blueshifted absorption component indicates that the velocity of the He-rich material is 1050$\pm$320 \kms. The feature detected in the first spectrum at $\sim$ 4600-4700\,\AA\ is now more prominent, and it shows a double-peaked profile. The red component is likely due to \Heii $\lambda$4686, while the blue component possibly arises from a blend of \Niii~$\lambda$4640 and \Ciii~$\lambda$4648. They are identified as flash-ionisation features, resembling those seen in other Type Ibn events, such as SNe 2010al \citep{Pastorello2015MNRAS.449.1921P}, 2019cj \citep{Wang2024A&A...691A.156W}, 2019uo \citep{Gangopadhyay2020ApJ...889..170G}, 2019wep \citep{Gangopadhyay2022ApJ...930..127G}, and 2023emq \citep{Pursiainen2023ApJ...959L..10P} (see further discussion in Sec.~\ref{subsec:SpectroComparison}). 
We note that the apparent emission bump at around 6560 \AA~is likely due to a blend of \Ha~and \Heii~$\lambda$6560 in these early spectra.

From $-$0.8 to $+$0.5 days after maximum light, the spectra are still dominated by blue continua with $T_{\rm BB}$ decreasing from $17,500 \pm 700~{\rm K}$ to $13,100 \pm 300~{\rm K}$.  The narrow P-Cygni profiles of \Hei~$\lambda$5876 become progressively more prominent with measured expansion velocities of 1050$-$1270 \kms. These P-Cygni features are likely produced in the He-rich CSM moving at a velocity slightly above 1000 \kms. The feature at 4600--4700\,\AA\ gradually becomes weaker during this time window,  disappearing at later phases.
The following three spectra, from $+$0.7 days ($T_{\rm BB} = 12,820 \pm 620~{\rm K}$) to $+$1.8 days ($T_{\rm BB} = 11,200 \pm 600~{\rm K}$), do not show significant evolution. The measured \Hei~$\lambda$5876\,\AA\ P-Cygni line velocities are approximately 1280 \kms~and 990 \kms, respectively.  
The only Balmer line detected in SN\,2024acyl is weak \Ha, which shows minor evolution in strength during these phases.

The following spectra, from $+$15.6 to $+$42.8 days, exhibit major changes. The continua are now much redder, with a temperature decreasing from $T_{\rm BB} = 8,300\pm300~{\rm K}$ to $T_{\rm BB} = 7,100\pm300~{\rm K}$ with a similar $\mathrm{log}(\sigma/10^{42}\,\mathrm{erg\,s^{-1}})\approx-3.9$. The emission components of \Hei~$\lambda$5876 dominate over the P-Cygni absorption starting from $+$15.6 days. The FWHM velocity of these broader \Hei~$\lambda$5876 emission lines, as obtained from a single Gaussian fit, is about 5800--6200 \kms. The broader profile observed for all lines indicates that the SN photosphere recedes with time from the CSM to the ejecta. As shown in the bottom of Fig.~\ref{fig:SN_spectra}, relatively broad features are identified in the blue region, including \Hei~$\lambda$3889, $\lambda$4471, $\lambda$4921, and $\lambda$5016, several \Feii~multiplets (e.g. \Feii~multiplet 42 lines at $\lambda\lambda\lambda$~4924, 5018, 5169) blended with \Hei~emission lines, and also \Mgi~$\lambda\lambda$4571, 5528  mostly in emission. In addition, \Hei~$\lambda$5876, $\lambda$6678, $\lambda$7065, $\lambda$7281 evolve significantly, becoming the most prominent emission features in the red spectral region ($\geq$\,5600\,\AA). \Ha~becomes more evident at late phases, blended with \Hei~$\lambda$6678, indicating the presence of hydrogen in the outer CSM. The near-infrarad (NIR) \Caii~triplet is weak in the $+$18.6 day NTT/EFOSC2 spectrum, while the emission at about 7300~\AA\ is likely a blend of [\Caii] $\lambda$$\lambda$7291, 7323 and \Hei~$\lambda$7281. We also tentatively identify \Oi~$\lambda\lambda\lambda$7772, 7774, 7775 lines in this spectrum, following  \citet{Pastorello2015MNRAS.454.4293P}. A relatively strong bump feature detected at 9000--9400\,\AA\ is tentatively identified as a blend including \Mgii~($\lambda$9218--9244). The late-time spectra of SN~2024acyl show an evident pseudocontinuum bluewards of $\sim$5600\,\AA. 
As suggested by \citet{Turatto1993MNRAS.262..128T}, \citet{Smith2012MNRAS.426.1905S}, and \citet{Stritzinger2012ApJ...756..173S}, it is likely due to a forest of narrow and intermediate-width Fe lines, as marked in the shaded region of Fig.~\ref{fig:SN_spectra}. The broad W-shape feature at 4600--5200\,\AA\ may also be attributed to Fe features blended with \Hei~lines. All the above features are frequently observed in late-time spectra of SNe Ibn.

\subsection{He~\textsc{i} line evolution}
\label{subsec:He_Lines}

\begin{figure*}
\begin{center}
\includegraphics[width=2\columnwidth]{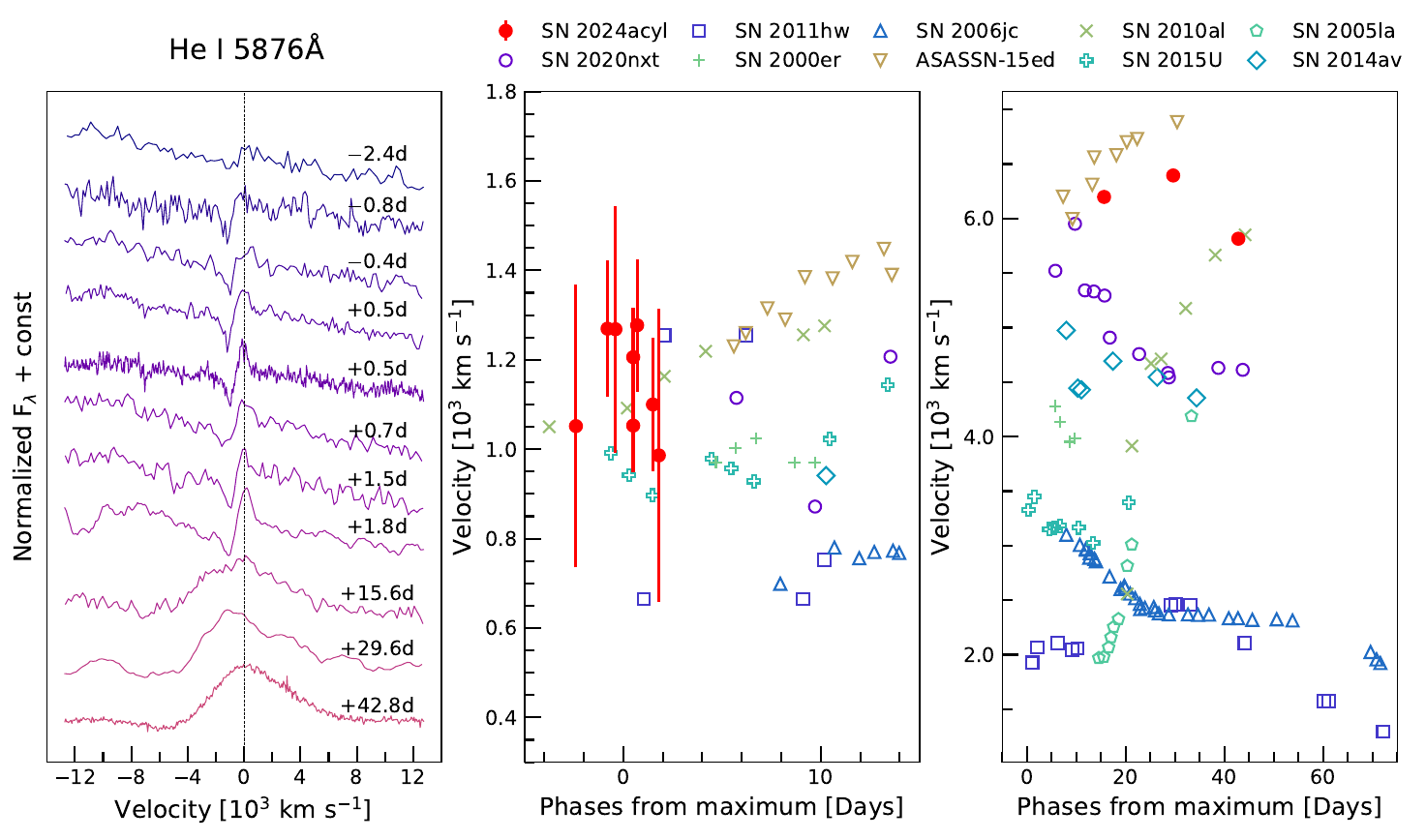}
\caption{Temporal evolution of the \ion{He}{i} $\lambda5876$ line. 
\textit{Left panel:} Line-profile evolution in the velocity space. The dashed vertical line marking marks the rest wavelength.
\textit{Middle panel:} Evolution of the velocities measured from the P-Cygni absorption minimum of the narrow \Hei component, formed in the unshocked CSM.
\textit{Right panel:} Evolution of the broader \ion{He}{i} emission components, reflecting the dynamics of the shocked gas. For clarity, the uncertainties are not shown in the plots, but they can reach values of up to  30\%. 
The comparison data for SNe~Ibn are from \citet{Pastorello2016MNRAS.456..853P} and \citet{Wang2025A&A...700A.156W}.}
\label{fig:HeVelocity}
\end{center}
\end{figure*}

In Fig.~\ref{fig:HeVelocity}, we illustrate the temporal evolution of the \ion{He}{i}~$\lambda5876$ line profile in SN~2024acyl (left panel).
The line strength increases with time, and two distinct kinematic components can be identified. 
The narrow \ion{He}{i} feature, with velocities of about 1000--1300~km~s$^{-1}$ as inferred from either the FWHM of the emission or the position of its weak P-Cygni absorption, arises from the slowly moving, unshocked He-rich CSM. 
A broader component, with a velocity of 5800--6400~km~s$^{-1}$, is detected in the later spectra. 
The coexistence of two components is consistent with an origin in two distinct regions, with the narrow P-Cygni features forming in the unperturbed CSM and the broader lines arising from the expanding ejecta. 
Similar evolution has been observed in the spectra of other SNe~Ibn, such as ASASSN-15ed \citep{Pastorello2015MNRAS.453.3649P} and SN~2010al \citep{Pastorello2015MNRAS.449.1921P}, where narrow lines dominate at early phases before broader ejecta signatures emerge. 
At early times, the photosphere lies within the dense CSM shell, above which the narrow lines form. 
These features are likely photoionised either by early ejecta--CSM interaction or by the initial shock breakout. 
As the shell recombines and becomes transparent, the underlying SN ejecta gradually dominate the spectra.

The middle panel of Fig.~\ref{fig:HeVelocity} shows the velocity evolution of the narrow \ion{He}{i} component. 
In most SNe~Ibn, these lines exhibit little or no change over time, consistent with emission from a quasistationary CSM shell. 
Typical velocities fall in the range 600--1500~km~s$^{-1}$, comparable to Wolf–Rayet wind speeds. 
For SN~2024acyl, the narrow-line velocity of $\sim$1100~km~s$^{-1}$ is similar to those measured in SNe~2020nxt, 2015U, and 2010al, while lower values ($<$\,800~km~s$^{-1}$) are found in SNe~2005la, 2011hw, and 2006jc. 
This spread in velocity likely reflects differences in progenitor wind properties such as terminal velocity and mass-loss rate.

The right-hand panel of Fig.~\ref{fig:HeVelocity} illustrates the velocity evolution of the broader \ion{He}{i} components. 
Unlike the narrow features, these display more pronounced temporal changes,
pointing to a diversity in the ejecta kinematics and in the density structure of the shocked CSM among the SNe Ibn of the comparison sample. 
For example, in SN~2006jc the intermediate-width \ion{He}{i} components narrowed from about 3100~km~s$^{-1}$ to 1700~km~s$^{-1}$ over four months, while in SN~2005la the velocities increased from $\sim$2000~km~s$^{-1}$ shortly after discovery to $\sim$4200~km~s$^{-1}$ within three weeks. 
However, the limited number of available spectra for SN~2024acyl prevents us from tracing a clear evolutionary trend.

\subsection{Comparison with Type Ibn SN spectra}
\label{subsec:SpectroComparison}

\begin{figure}[htbp]
\includegraphics[width=1\columnwidth]{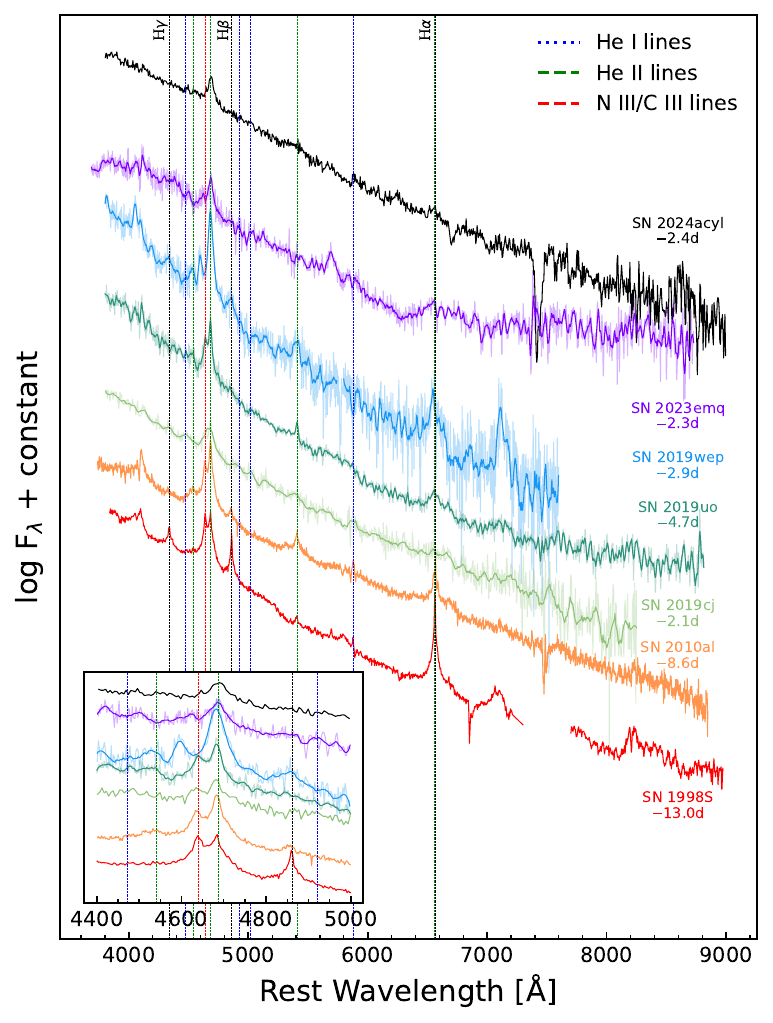}
\caption{Comparisons of the spectra of SN~2024acyl with the Type IIn SN~1998S and other Type Ibn events (SN~2010al, SN~2019cj, SN~2019uo, SN~2019wep and SN~2023emq) at their very early phases. The inset shows a close-up view of the region between 4400\,\AA\ and 5000\,\AA\ with prominent flash-ionisation features. The phases marked on the right side of each SN spectrum are with respect to the epoch of their maximum light. Spectra with a low S/N were binned with 20~\AA\ in each bin. The original (unbinned) spectra are displayed in lighter colours behind.}
\label{fig:spectra_early}
\end{figure}

\begin{figure*}
    \includegraphics[width=1\linewidth]{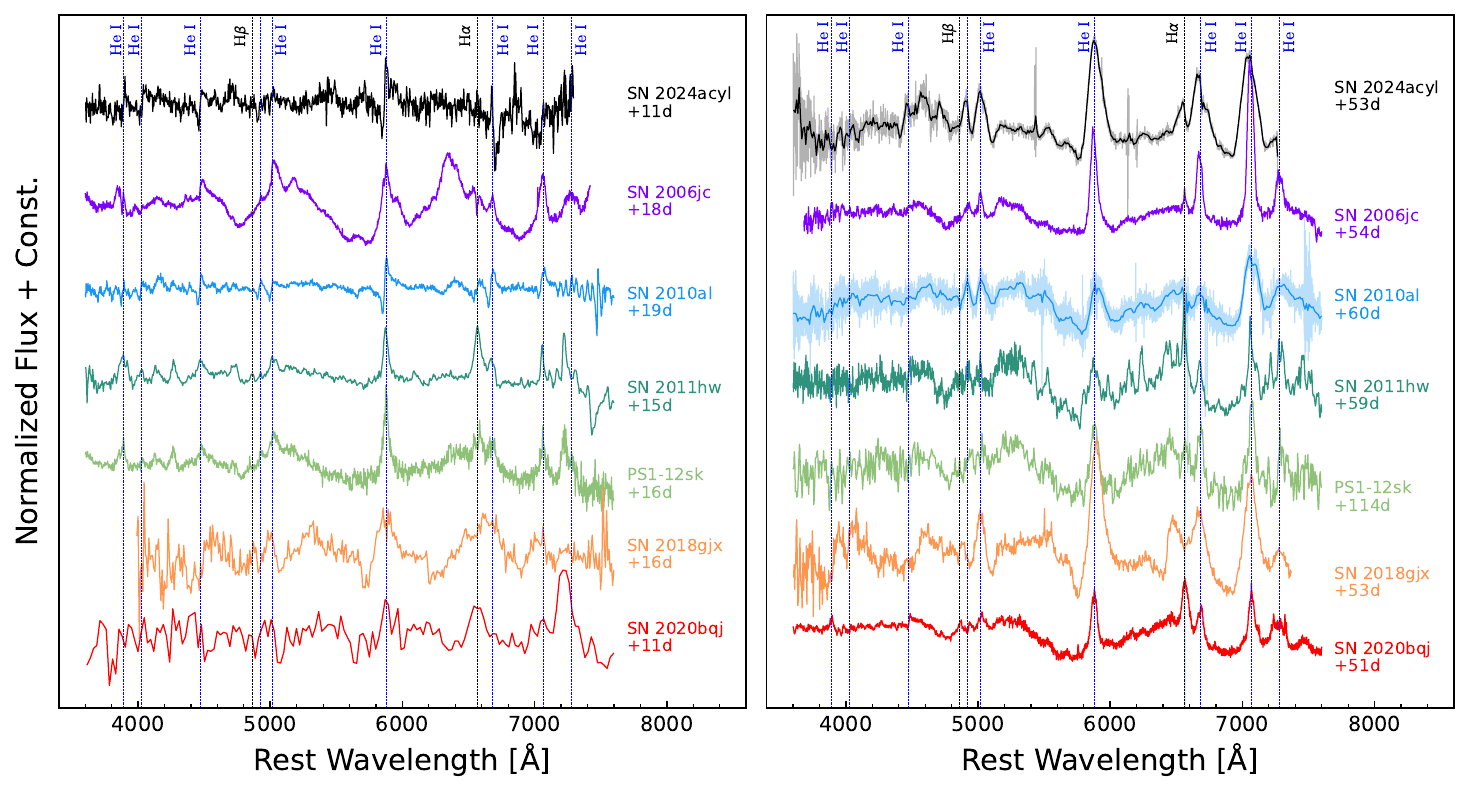}
    \caption{Comparisons of the spectra of SN~2024acyl at different phases with those of the transitional Type IIb/Ibn event SN~2018gjx and several Type Ibn events with H signatures, such as SNe~2006jc, 2010al, 2011hw, 2020bqj, and PS1-12sk. {\it Left panel:} Spectra obtained at around the time of maximum light ($\sim10-20 \, \mathrm{days}$). {\it Right panel:} Late-time spectra ($\sim50-120 \, \mathrm{days}$). The key spectral lines (H and He) are marked with coloured dashed lines. The phases marked on the right side of each spectrum are with respect to the epoch of their maximum light. Spectra with low S/N have been binned with 20~\AA; the original (unbinned) spectra are displayed in lighter colours behind. All the phases marked in the figure are related to the (approximate) explosion epoch.}
    \label{fig:specta_PeakLate}
\end{figure*}

Figure~\ref{fig:spectra_early} compares the early-time spectrum of SN~2024acyl with those of several Type Ibn events (SNe~2010al, 2019cj, 2019uo, 2019wep, and 2023emq) as well as the Type IIn SN~1998S. All spectra were obtained within a few days prior to the epoch of the maximum light. The inset highlights the region between 4400 and 5000~\AA, where flash-ionisation features are most prominent. SN~2024acyl shows remarkable resemblance to other events, particularly in the simultaneous presence of flash-ionisation signatures of He~\textsc{ii} and prominent N/C lines. The detection of high-ionisation transitions, such as \ion{He}{ii}~$\lambda4686$ and \ion{N}{iii}/\ion{C}{iii}~$\lambda\lambda4640,4650$, indicates the action of an external photoionisation source on the dense CSM, most likely arising from a shock breakout or the onset of ejecta--CSM interaction. Such transient flash-ionisation signatures are observed only in a subset of SNe Ibn \citep[e.g.][]{Pastorello2016MNRAS.456..853P, Gangopadhyay2020ApJ...889..170G, Gangopadhyay2022ApJ...930..127G, Pursiainen2023ApJ...959L..10P, Wang2024A&A...691A.156W}, but are well documented in the early spectra of many CC Type II SNe \citep{Fassia2001MNRAS.325..907F, Gal-Yam2014Natur.509..471G, Bostroem2023ApJ...956L...5B, Bruch2023ApJ...952..119B, Zhang2023SciBu..68.2548Z, Zhang2024ApJ...970L..18Z, Jacobson2024ApJ...972..177J}. 
Furthermore, the early detection of nitrogen lines in SN~2024acyl points to the presence of CNO-processed material in the progenitor wind, thereby providing constraints on its pre-SN evolutionary state.

Figure~\ref{fig:specta_PeakLate} illustrates the spectral evolution of SN~2024acyl compared with a diverse sample of Type Ibn and IIb/Ibn events at both intermediate ($+10$ to $+20$~d; left panel) and later epochs ($+50$ to $+100$~d; right panel). The phases are related to the estimated explosion epoch, in order to make the comparisons more resonable. Around maximum light, SN~2024acyl exhibits pronounced P-Cygni profiles in He~\textsc{i} $\lambda\lambda4471, 5876, 7065$, closely resembling SN 2010al at comparable phases \citep{Pastorello2015MNRAS.449.1921P}. Weak but clearly detectable Balmer emission lines are also present, placing SN~2024acyl within the H-bearing subset of Type Ibn events.
At later phases, SN~2024acyl still maintains prominent He~\textsc{i} emission lines with increasingly broader widths, consistent with other Type Ibn SNe. The persistent helium features, combined with the enhanced Balmer components, indicate that the CSM is primarily helium-rich, with hydrogen confined to the outermost layers. 
Taken together, these spectral comparisons confirm that the ejecta of SN~2024acyl interact with a dense, helium-dominated CSM that contains a residual amount of hydrogen. The spectroscopic diversity of SNe Ibn likely depends on the variety of their progenitor mass-loss histories, wind compositions, or how the stellar components in a binary system interact in the final stages before the core collapse.

\subsection{Spectral modelling}
\label{subsec:SpectroModelling}

\begin{figure}[htp]
\includegraphics[width=1\columnwidth]{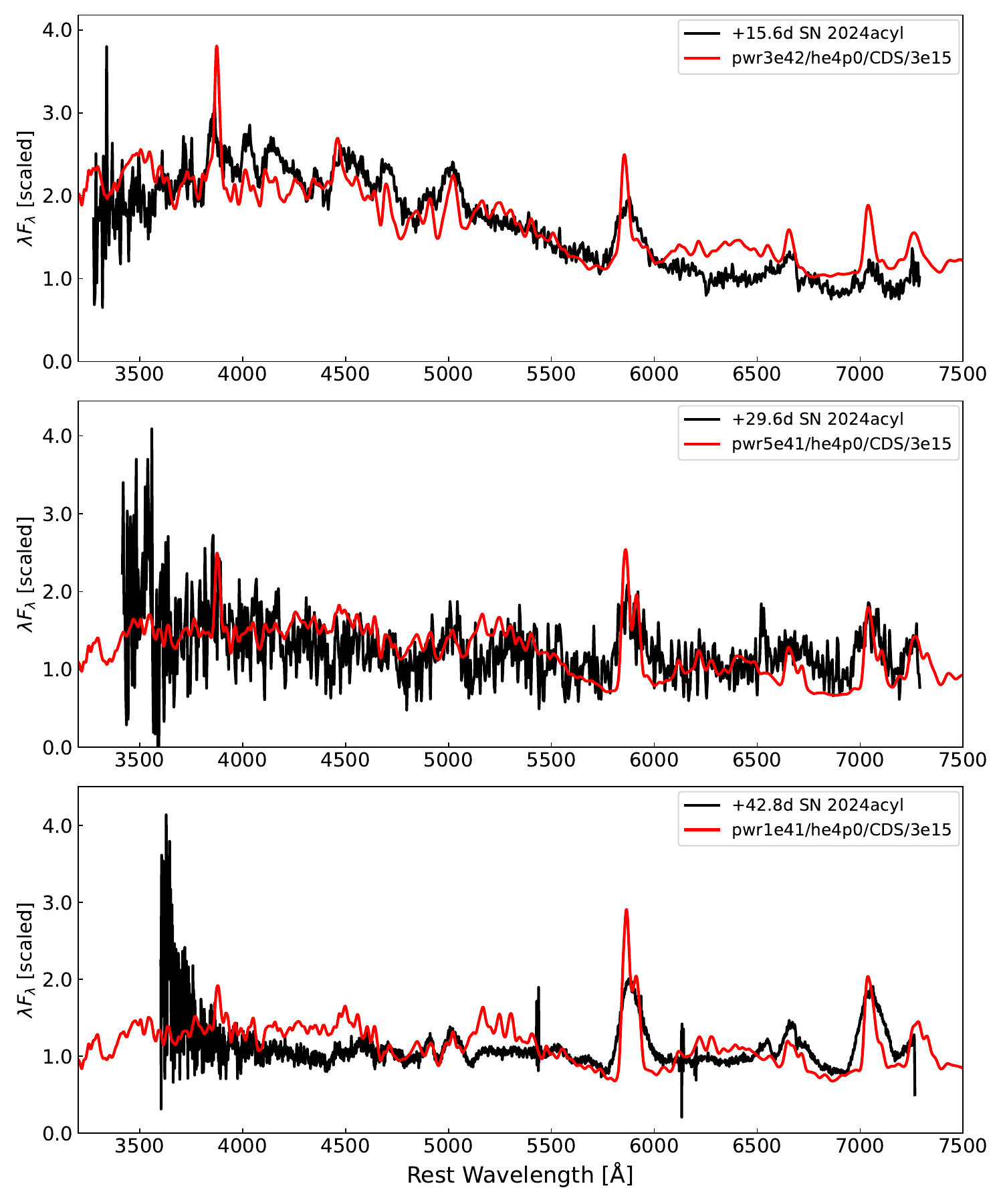}
\caption{Comparison of synthetic spectra from the \texttt{he4p0} model with the observed spectra of SN 2024acyl obtained after the $o$-band maximum light. No smoothing has been applied to either the observed or the synthetic spectra. The synthetic spectra are based on simulations by \citet{Dessart2022A&A...658A.130D} and \citet{Wang2024MNRAS.530.3906W}, as well as on newly computed models incorporating updated parameters. The model spectra have been scaled to match the resolution of the observed spectra. } 
\label{fig:spectra_model}
\end{figure}

To investigate the ejecta properties and the progenitor system of SN~2024acyl, we compared a subset of the observed spectra with a series of non-local thermodynamic equilibrium radiative-transfer models computed with \textsc{CMFGEN} \citep[][]{Hillier2012MNRAS.424..252H}. These include both the models of \citet{Dessart2022A&A...658A.130D} and additional tailored simulations with modified input parameters (Dessart, priv. comm.). The adopted configurations involve the interaction of moderately energetic ($\sim 10^{50}$~erg), low-mass ($\lesssim 1~M_\odot$) ejecta with a slowly expanding, helium-rich circumstellar envelope of a roughly comparable mass, which is compatible with our \texttt{MOSFiT} results. Under this condition, the analysis of the low-energy output can be simplified by concentrating on the cool dense shell (CDS), which develops as a compact, thin layer during the interaction and becomes the dominant source of radiation at later phases.

In this steady-state approximation, the hydrodynamical evolution is not explicitly modelled, and the CDS is represented as a chemically mixed layer described by a Gaussian density distribution centred at 2000~km~s$^{-1}$ and scaled to recover the total ejecta mass. Energy originating from both RD and residual interaction is deposited non-thermally within the CDS, enabling a consistent treatment of its ionisation and excitation conditions. Although these simulations are not designed to reproduce individual SNe in detail, they provide a valuable tool to investigate the spectral diversity and the underlying parameter degeneracies. \citet{Dessart2022A&A...658A.130D} further demonstrated that similar spectral morphologies can arise from different combinations of CDS mass, radial extent, and input power.

The persistent presence of \ion{He}{i} lines throughout the spectral evolution indicates a progenitor dominated by a helium-rich composition. Therefore, following the methodology of \citet{Wang2024MNRAS.530.3906W,Wang2025A&A...700A.156W}, we adopted scenarios with a helium-star progenitor. According to the helium-star evolutionary tracks of \citet{Woosley2019ApJ...878...49W}, progenitors with zero-age `He' main-sequence masses of 3--4~$M_\odot$ (e.g. models \texttt{he3} and \texttt{he4}) are consistent with the observed properties of SNe~Ibn. In this work, we made use of model \texttt{he4}, characterised by  $M_{\mathrm{preSN}} = 3.16\ \msun$ and total ejecta mass  1.62~$M_\odot$, comprising 0.92~$M_\odot$ of helium, 0.31~$M_\odot$ of oxygen, 0.03~$M_\odot$ of magnesium, and 0.0014~$M_\odot$ of calcium, and assuming a solar metallicity, which indicate a low-mass progenitor scenario. Figure~\ref{fig:spectra_model} presents synthetic spectra from model \texttt{he4p0} compared with observations of SN~2024acyl at multiple epochs after maximum light. 
We present the model comparisons of the spectra in $+$ 15.6 days, $+$ 29.6 days, and $+$ 42.8 days since the $o$-band maximum. To highlight the role of luminosity evolution, only the input power changes with time. Thus, we adopt the \texttt{he4p0} models with $3\times10^{42}\,\mathrm{erg\,s^{-1}}$, $5\times10^{41}\,\mathrm{erg\,s^{-1}}$, and $1\times10^{41}\,\mathrm{erg\,s^{-1}}$ luminosity correspondingly, while the CDS radius ($3\times10^{15}$~cm) and velocity (2000~km~s$^{-1}$) were kept constant. The adopted power values are broadly consistent with the observed bolometric light curve (see Fig.~\ref{fig:quasibolom}).

At +15.6~d, the assumption of a narrow and homogeneous dense shell is unlikely to be valid. Relative to SN~2024acyl, the models systematically overpredict the strengths of the \ion{He}{i} lines
(see the top panel of Fig.~\ref{fig:spectra_model}). At +29.6~d and +42.8~d, the \Hei profiles predicted by the models exhibit a blue-red asymmetry and a central absorption dip (see Fig.~\ref{fig:spectra_model}, middle and bottom panels). However, the observations of SN~2024acyl reveal a \Hei line with a single, rounded peak profile. This suggests that the dense shell formed during the interaction is highly clumped.

The most striking difference between SN~2024acyl and the synthetic spectra is that some modelled features are narrower than those observed. This inconsistency could be alleviated by adopting higher CDS velocities in tailored models, which would broaden the lines while preserving the overall spectral morphology. Despite such local mismatches, the overall spectral evolution is well reproduced, providing strong support for the interpretation that the emission of SN~2024acyl is powered by ejecta--CSM interaction.

\section{Discussion} 
\label{sec:SummaryDiscussion}

In the above sections, we have presented the photometric and spectroscopic analyses of SN~2024acyl. We now piece the analyses together to place SN~2024acyl in the context of SNe Ibn, attempting to study its spectrophotometric properties and shed light on their progenitor systems. We summarise our findings of SN~2024acyl in the following.

\begin{itemize}
    \item The rise time of the light curve to maximum brightness is about 10.6 days, which is within the observed range of the Type Ibn samples of \citet{Hosseinzadeh2017ApJ...836..158H} and \citet{Wang2024A&A...691A.156W,Wang2025A&A...700A.156W}, $\sim 2$--20 days, although slightly longer than typical SNe Ibn \citep[$\sim$7 days; see Fig. 11 of][]{Wang2025A&A...700A.156W}, suggesting the need for a relatively high mass-loss rate to reproduce the early light curve. 
    \item The SN is relatively sub-luminous ($M_{o} = -17.58 \pm 0.15$ mag)  with respect to SNe Ibn \citep[$M_r \approx -19~\mathrm{mag}$;][]{Pastorello2016MNRAS.456..853P,Hosseinzadeh2017ApJ...836..158H,Wang2025A&A...700A.156W}. The estimated peak `optical' and `UV+Optical' luminosities are ($3.5\pm0.8) \times 10^{42}$ erg s$^{-1}$ and ($6.7\pm0.4) \times 10^{42}$ erg s$^{-1}$, respectively. The corresponding radiated energies are $(5.0\pm0.4) \times 10^{48}$ erg and $(8.5\pm0.6) \times 10^{48}$ erg, respectively. 
    \item The post-peak light-curve decline shows a fast and almost linear trend, with a decline rate of $\gamma_{0-60}($V$) = 0.097 \pm 0.002$ $\mathrm{mag \, day^{-1}}$ during its post-peak evolution. The fast decline in the light curves is attributed to the much lower ejecta mass and optical depth, resulting in the rapid release of stored radiative energy in a short time \citep{Dessart2024arXiv240504259D}. 
    \item We compared the $B-V$ and $r-i$ colour curves of SN~2024acyl with a number of SNe~Ibn, which generally exhibit some heterogeneity in their observed colour evolution. The substantial diversity in colour evolution among SNe~Ibn may reflect the different physics that govern the colour curves of these events.
    \item We performed multi-band light-curve fits using the \texttt{MOSFiT} code and adopted the RD+CSI model to constrain the properties both for the radioactive power and CSM. The posterior distributions of the parameters are well-converged, suggesting a relatively low ejecta mass of $0.49^{+0.11}_{-0.09}\, \msun$ and a correspondingly low kinetic energy of $0.06^{+0.01}_{-0.01} \times 10^{51} \, \mathrm{erg}$. The derived CSM properties are consistent with other Type Ibn SNe, with $M_{\mathrm{CSM}}=0.51^{+0.05}_{-0.04} \, \msun$ and an inner radius of $17.8_{-3.0}^{+3.6} \, \mathrm{AU}$. The $^{56}$Ni mass of SN~2024acyl is $0.018_{-0.005}^{+0.009}\,\msun$, consistent with the values inferred from other Type Ibn SNe \citep[e.g.][]{Gangopadhyay2020ApJ...889..170G, Kool2021A&A...652A.136K, Maeda2022ApJ...927...25M, Wang2025A&A...700A.156W}. The light-curve fitting results indicate that SN~2024acyl is a less energetic event with small ejecta mass that occurred in a CSM environment typical for Type Ibn SNe. 
    \item The spectral evolution can be divided into three distinct phases. The early spectra from $-$3.7 to $+$0.5 days exhibit relatively slow spectral evolution, with hot blue continua and the narrow P-Cygni profiles of the \Hei~lines. In addition, these early spectra are characterised by the prominent flash-ionisation lines of \Ciii, \Niii, and \Heii, which are occasionally detected in SNe Ibn. From $+$0.7 to $+$1.8 days, the flash-ionisation signatures have completely disappeared, while the narrow \Hei~lines with P-Cygni profiles show a modest evolution. Later spectra after $+$15.6 days show significant changes with prominent and broad \Hei~emission lines, along with the appearance of \Feii, \Caii, \Mgi, and \Oi~lines. Additionally, we find a flux drop in the pseudocontinuum bluewards of $\sim$5600\,\AA, likely from a forest of \Feii~lines. The \Ha~feature is detected in almost all spectra of SN~2024acyl, faint at early times but becoming prominent at late phases. The \Ha~feature detected in transitional SNe Ibn (in particular at late times) indicates the presence of a residual amount of H in the outer CSM.
\end{itemize}

\label{sec:physics_2024acyl}

\subsection{A possible massive Wolf-Rayet-like progenitor}

In general, the evolution of SN~2024acyl is similar to that of typical SNe Ibn in its photometric and spectroscopic properties (see details in Sec. \ref{sec:photometry} and Sec. \ref{sec:spectroscopy}). However, the main difference between normal SNe Ibn and SN~2024acyl is the existence of weak H emission lines. Such a feature suggests SN~2024acyl is a new case of a transitioning SN Ibn/IIn. Similar transitional spectroscopic features are only occasionally observed \citep[e.g. SNe 2005la, 2010al, 2011hw, 2021foa; see][]{Pastorello2008MNRAS.389..131P,Smith2012MNRAS.426.1905S,Pastorello2015MNRAS.449.1921P,Reguitti2022A&A...662L..10R,Farias2024ApJ...977..152F,Gangopadhyay2025MNRAS.537.2898G}. However, the discovery of these transitional interacting SNe suggests the existence of a continuum in the properties (such as mass-loss history) and progenitor types, between at least some SNe IIn and SNe Ibn. The intensity of the H lines gradually decreases from the H-dominated SNe IIn, through SN\,2009ip-like (with strong H plus weak He), SN\,2021foa (with strong H and He), SN\,2006jc (with weak H and strong He), to SNe Ibn \citep[only showing narrow~\Hei; see e.g. Fig.~5 of][]{Pastorello2025A&A...701A..32P}, with SN~2024acyl sitting somewhere in the middle. In this context, SN~2024acyl and other transitional SN Ibn/IIn events have been proposed to result from the explosion of massive stars that were transitioning from the LBV to the WR stages \citep[][]{Pastorello2008MNRAS.389..131P,Smith2012MNRAS.426.1905S,Pastorello2015MNRAS.449.1921P,Reguitti2022A&A...662L..10R}. SN~2024acyl has somewhat hybrid properties between SNe Ibn and IIn. In particular, the late-time spectra suggest that the outer envelope of the SN\,2024acyl progenitor had residual H at the time of explosion. However, the H/He line intensity ratio indicates that the progenitor of SN\,2024acyl was much more H-stripped than SNe\,2011hw, 2020bqj, and 2021foa. Therefore, the progenitor of SN\,2024acyl could be a late-type WR star with hydrogen, or even an Ofpe/WN9 star.\footnote{The Ofpe/WN9 stars, also known as `slash stars', were introduced by \citet{Walborn1977ApJ...215...53W,Walborn1982ApJ...256..452W} These objects are characterised by the presence of strong low-ionisation emission features (\Nii, \Hei) along with the normal high-ionisation Of-features (\Niii, \Heii) in their spectra. They exhibit spectroscopic properties that are intermediate between those of Of and WN stars, indicating that they are transition objects between Of and WR stars \citep[e.g.][]{Conti1975MSRSL...9..193C,Bianchi2004ApJ...601..228B}.}.

The mass-loss rate of the progenitor can be estimated from the CSM properties and stellar wind velocity. However, quantifying the mass loss via stellar winds is complex, and owing to the incomplete dataset lacking X-ray observations \citep[with which the mass-loss history can be accurately derived;][]{Pellegrino2024ApJ...977....2P} and the shell-like density profile of CSM \citep[in which the velocity of the wind is not steady;][]{Ben-Ami2023ApJ...946...30B}, we can only constrain the order of magnitude of the mass-loss rate. Thus, taking the relation from \cite{Chatzopoulos2012ApJ...746..121C}, the mass-loss rate can be expressed as
\begin{equation}
\dot M(r) = 4 \pi r^{2-s} \rho_{\mathrm{CSM}}\ R_0^s\ v_{\mathrm{wind}}\, .
\end{equation}
We adopted the best-fit posterior values: an inner CSM radius of $r=R_0=17.8_{-3.0}^{+3.6} \, \mathrm{AU}$ and a CSM density of $\rho_{\mathrm{CSM}} = (8.3_{-1.2}^{+2.7})\times10^{-12} \, \mathrm{g\,cm^{-3}}$. Assuming a shell-like density profile ($s=0$), the mass-loss rate was calculated to be $\dot M \approx 11.7\,(v_{\mathrm{wind}}/(1000\,\mathrm{km\,s^{-1}}))\,\msun\,\mathrm{yr^{-1}}$. For a typical WR star wind velocity of $v_{\rm{w}}\approx1000\,\mathrm{km\,s^{-1}}$ \citep[][]{Chevalier2006ApJ...651..381C}, the mass-loss rate is thus $11.7 \,\msun\,\mathrm{yr^{-1}}$. This rate is consistent in magnitude with the range found by \cite{Ben-Ami2023ApJ...946...30B}. Alternatively, following the approach of \cite{Maeda2022ApJ...927...25M} for increasing winds from a WR-like progenitor (corresponding to $s=3$), the mass-loss rate is $\dot M \approx 5.5\,\msun\,\mathrm{yr^{-1}}$. Both estimates are exceptionally high, significantly exceeding the typical range of $10^{-3}$--$10^{0}\,\msun\,\mathrm{yr^{-1}}$ for stellar winds \citep{Nyholm2017A&A...605A...6N, Wang2020ApJ...900...83W}. These results suggest that a steady mass-loss scenario is unlikely, and an enhanced mass-loss episode a few years before the explosion should be considered. 
Therefore, our analysis does not rule out the possibility that the progenitor of SN~2024acyl was compatible with a WR-like star experiencing enhanced mass loss shortly before core collapse.
As further support, the CSM velocity (990--1280 \kms) measured for SN~2024acyl 
is significantly faster than LBV winds \citep{Smith2017hsn..book..403S}.

Additionally, for a CSM shell expanding at $v\approx 1000\,\mathrm{km\,s^{-1}}$, the travel time to a radius of $R\approx 17.8\,\mathrm{AU}$ is about 30 days. We also estimate the duration of the eruptive mass-loss event to be $\sim 6$ days. The kinetic energy released during this event is substantial, around $5\times10^{48}\,\mathrm{erg}$, which is not negligible compared to the kinetic energy of SN~2024acyl. However, no optical emission was detected in the ATLAS data in the 20--40 days prior to the explosion. Furthermore, we did not find any pre-SN detection in Pan-STARRS archival data, corresponding to a 3$\sigma$ limit of $m \gtrsim 22\,\mathrm{mag}$. Given the limiting magnitude of the ATLAS survey ($\sim$20 mag in the $o$ band), this non-detection implies that only luminous pre-explosion eruptions would have been detectable. At this distance, the detection limit corresponds to an absolute magnitude brighter than about $-15$ mag. For comparison, the pre-explosion outburst of SN~2006jc was detected at roughly $-14$ mag \citep{Pastorello2007Natur.447..829P}. This interpretation also suggests that the mass-loss eruption could have been fainter, yet still possibly  producing dense and optically thick ejecta. Moreover, the later observation of flash-ionised \ion{C}{III} and \ion{N}{III} lines implies an extended progenitor, as suggested by \citet{Blinnikov2003fthp.conf...23B}, likely resulting from the intense mass-loss process. Therefore, the late-type WR progenitor is likely a potential interpretation for SN\,2024acyl. 

We remark that a black hole might be formed through fallback accretion with no or weak SN explosion, and thus no or little $^{56}$Ni will be ejected and the ejecta mass will also be low \citep[e.g.][]{Woosley1995ApJS..101..181W, Zampieri1998ApJ...502L.149Z, Maeda2007ApJ...666.1069M, Moriya2010ApJ...719.1445M}. This scenario is consistent with the constraints on physical parameters derived from the light-curve modelling of SN\,2024acyl --- lower ejecta mass of $M_{\mathrm{ej}}$ of $0.49^{+0.11}_{-0.09} \, \msun$~and lower $^{56}\mathrm{Ni}$ mass of $M_{\mathrm{Ni}} = 0.018 \, \msun$ (see details in Sec. \ref{Sec:MOSFiT}). 
However, the fallback accretion model predicts a lower ejecta velocity, a slower light-curve decline rate compared to typical Type Ibn SN samples, and a possible optical afterglow of a gamma-ray burst (GRB) \citep{Moriya2010ApJ...719.1445M}. In contrast, our spectroscopic analysis of SN~2024acyl reveals a relatively high ejecta velocity. Moreover, its light-curve decline rate is consistent with other SN Ibn samples, and its SEDs can be well described by a single black-body model. These observed properties challenge the features predicted by the fallback accretion scenario.
Furthermore, another prediction of the fallback-enforced explosion model is a chemical composition rich in oxygen, carbon, and magnesium, but poor in iron \citep{Moriya2010ApJ...719.1445M}. This appears to contradict our spectra of SN~2024acyl, which exhibit prominent iron features, as do other typical Type Ibn events and stripped-envelope SN samples (see Fig.~\ref{fig:specta_PeakLate}), but lack a clear detection of oxygen lines, particularly the [\Oi] $\lambda\lambda$6300, 6364\,\AA\ doublet. However, the non-detection of these [\Oi] lines is not conclusive. Their intensity is highly sensitive to the ejecta density \citep{Valenti2009Natur.459..674V}, and our spectrum at +42.8 days was likely obtained before the ejecta became optically thin to these forbidden lines. Therefore, the fallback accretion scenario cannot be definitively ruled out on the basis  of the current spectral evidence. 

\subsection{SN~2018gjx-like: Type IIb event that exploded in He-rich CSM}
\begin{figure}
    \centering
    \includegraphics[width=1.0\linewidth]{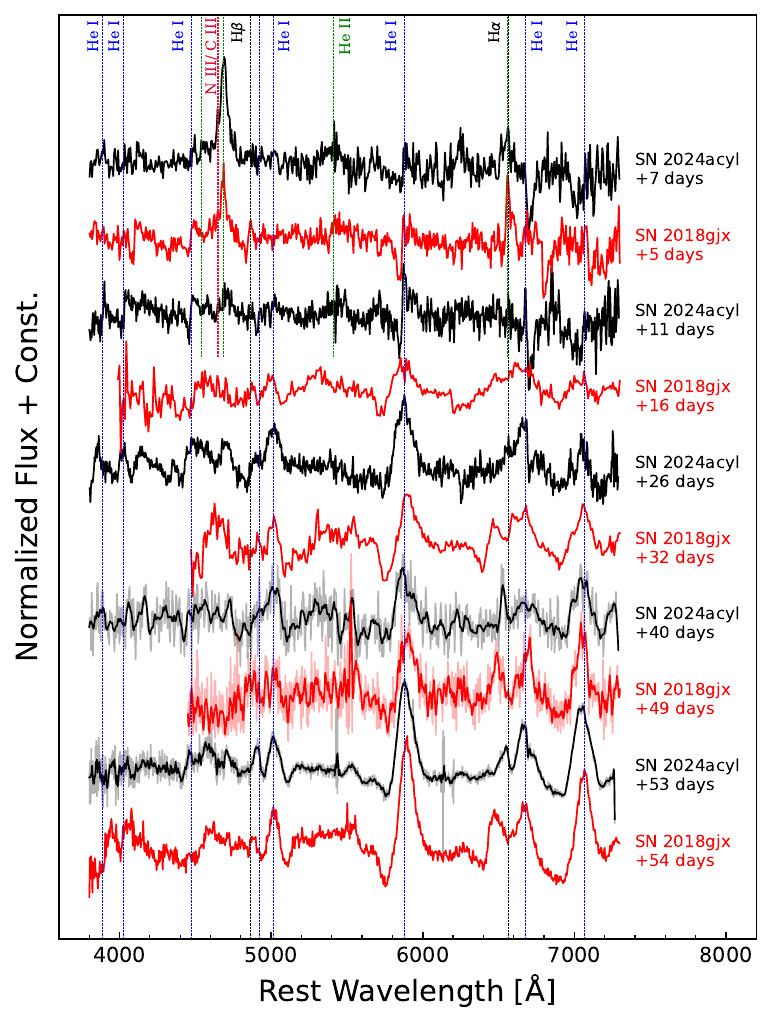}
    \caption{Comparisons of the spectra of SN~2024acyl with SN~2018gjx at multiple epochs. The SN~2024acyl spectra are in red; SN~2018gjx spectra are in black. Spectra with a low S/N were binned. The original (unbinned) spectra are displayed in lighter colours behind. All the phases marked in the figure are from the (approximate) explosion epoch.}
    \label{fig:24acyl_speccmp_gjx}
\end{figure}

We compared SN~2024acyl with SN~2018gjx, a Type IIb SN that exploded in He-rich CSM \citep{Prentice2020MNRAS.499.1450P}, from early to the late phases as shown in Fig.~\ref{fig:24acyl_speccmp_gjx}. During the early phase ($t\lesssim11$~days), both SN~2024acyl and SN~2018gjx display significant signatures of interaction with the CSM. Narrow emission lines from highly ionised species such as \ion{N}{III} and \ion{C}{III}, along with \ion{He}{II}, appear in the spectra. These features are also observed in other strongly interacting Type IIb events, such as SN~2017ckj \citep{Li2025A&A}. Furthermore, a narrow \ion{He}{I} P~Cygni profile at 5876~\AA\ is visible in the spectra of both objects. However, we note a difference in the intensity of the H$\alpha$ emission. We do not detect prominent H$\alpha$ emission in the early phase of SN~2024acyl compared with SN~2018gjx. This inconsistency indicates that the progenitor of SN~2024acyl is H-poor. In this phase, the photosphere is located at the outer boundary of the optically thick CSM \citep{Chatzopoulos2012ApJ...746..121C}.

During the phases $15\lesssim t\lesssim40$~days (see the fourth to seventh spectra in Fig.~\ref{fig:24acyl_speccmp_gjx}), broad lines begin to appear. In SN~2018gjx, the P-Cygni profiles of \ion{He}{i} become broad and prominent, originating from the expanding opaque ejecta \citep{Prentice2020MNRAS.499.1450P}. This indicates that the ejecta have extended beyond the CSM and the photosphere is currently located near the outer boundary of the ejecta. Additionally, features of Type IIb SNe, such as weak H$\alpha$ absorption, are present in SN~2018gjx. However, we do not observe these features in SN~2024acyl during  comparable phases. In the spectrum at $+$26~days from explosion, there are no prominent broad P~Cygni profiles. This discrepancy may be attributed to differences in CSM properties. For SN~2024acyl, the CSM is optically thick ($\tau=\int\kappa\rho\,\mathrm{d}r\gtrsim100$), and $M_{\mathrm{CSM}}$ is comparable to $M_{\mathrm{ej}}$. Consequently, the formation of a CDS obscures the spectroscopic features of the ejecta \citep{DessartMNRAS2015, Dessart2022A&A...658A.130D}. Furthermore, given that the radioactive power in SN~2024acyl is much weaker than that of SN~2018gjx, the luminosity of SN~2024acyl is likely dominated by CSM interaction. Unlike SN~2018gjx, SN~2024acyl exhibits strong signatures of CSM interaction throughout its evolution.

At $t\gtrsim40$~days (see the eighth to tenth spectra in Fig.~\ref{fig:24acyl_speccmp_gjx}), broad emission lines of \ion{He}{i} appear in the spectra of both SNe. This indicates that the photosphere is receding as the ejecta become optically thin (where $\tau\propto t^{-2}$). Consequently, the interaction with the CSM becomes visible again, resulting in broad emission lines, a feature commonly observed in Type Ibn SNe. Furthermore, the spectra of both SN~2024acyl and SN~2018gjx display potential P~Cygni absorption features associated with \ion{He}{i}~$\lambda5876$. For SN~2024acyl, the velocity estimated from this possible P~Cygni absorption is $v_{5876}=4690^{+550}_{-540}\,\mathrm{km\,s^{-1}}$,  significantly higher than both the wind velocity and the initial CSM velocity ($\sim1000\,\mathrm{km\,s^{-1}}$). This suggests that the absorption profile originates from the expanding ejecta or the CDS. Since apparent absorption features can sometimes mimic dips in the pseudocontinuum caused by metal lines \citep[e.g.][]{Pastorello2015MNRAS.449.1921P, Kool2021A&A...652A.136K}, it is necessary to confirm that these features originate from optically thick ejecta. Following the method of \citet{Prentice2020MNRAS.499.1450P}, we estimate the velocity of the possible absorption feature of \ion{He}{i}~$\lambda6678$. We measure a velocity of $v_{\mathrm{6678}}=4460_{-180}^{+270}\,\mathrm{km\,s^{-1}}$. This value is consistent with the estimate obtained from \ion{He}{i}~$\lambda5876$, reinforcing the interpretation that these absorption features likely originate from the expanding ejecta and/or CDS. The profile of the \ion{He}{i}~$\lambda5876$ broad emission line also shows a profile similar to that of SN~2018gjx, which is symmetric in the late phases, indicating that there was no prominent dust formation in SN~2024acyl similar to that in SN~2018gjx. Additionally, an asymmetric or toroidal CSM geometry \citep{SmithMNRAS2015, Prentice2020MNRAS.499.1450P} may explain the visibility of these features, as ejecta signatures can remain observable through regions of lower optical depth.

Although there are several similarities between SN~2018gjx and SN~2024acyl, the relatively weak H$\alpha$ emission line in the early phases and the strong signature of CSM interaction throughout its evolution make it difficult to classify SN~2024acyl as a Type IIb SN exploding in He-rich CSM. Compared to SN~2018gjx, SN~2024acyl is more likely an H-poor event. The signature of H$\alpha$ in SN~2024acyl is weak in early phases compared with SN~2018gjx. In the late phases of SN~2024acyl, H$\alpha$ gradually becomes prominent, but present as a pure narrow emission line without any P~Cygni profile. This may be due to a small amount of residual hydrogen being excited during the late phases in an optically thin region, which is similar to most of the SNe Ibn such as SN~2006jc and SN~2020bqj. The discrepancy of CSM interaction strength can be explained either by inherently strong interaction, as previously discussed, or by a toroidal CSM geometry viewed edge-on. Also, since the case for a WR star as the progenitor of SN~2024acyl is less favoured in the previous discussion, it is also difficult to explain the progenitor of SN 2024acyl within the framework of a Type IIb event. Although these features of SN~2024acyl share similar physical mechanisms with SN~2018gjx, they are inconsistent with an SN~IIb classification.

\subsection{A possible low-mass progenitor}

\begin{figure}
    \centering
    \includegraphics[width=1.0\linewidth]{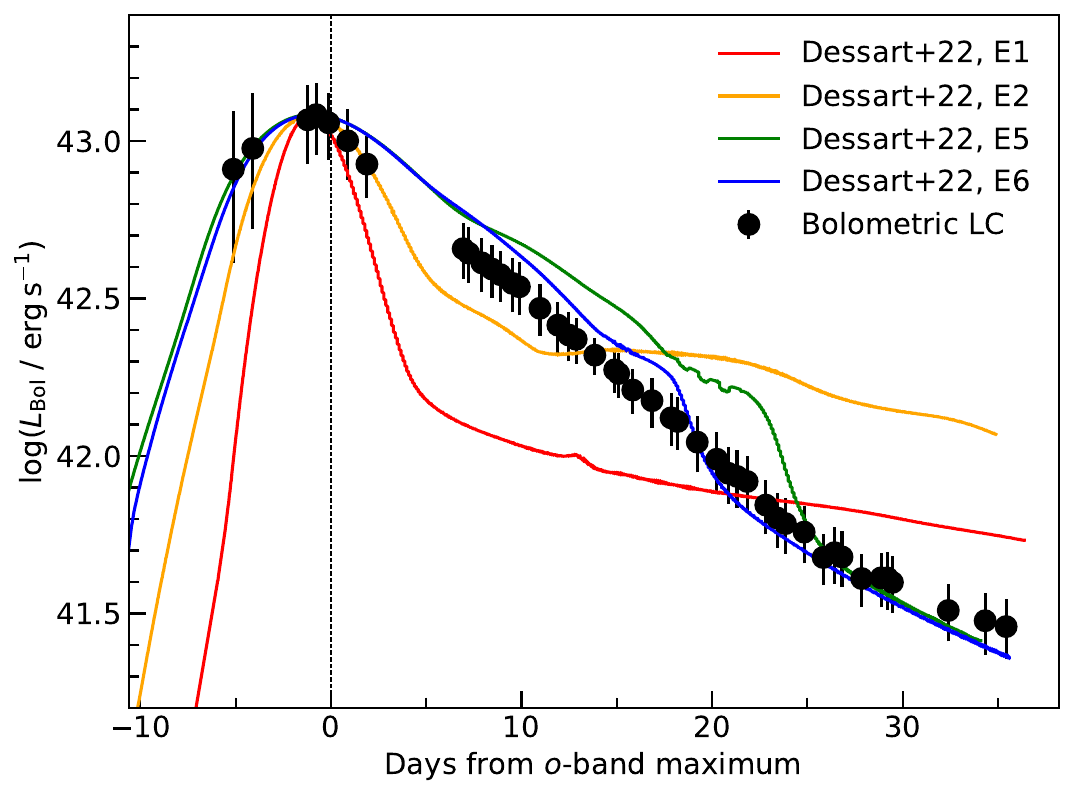}
    \caption{Bolometric light curve of SN~2024acyl compared with several selected interaction models from \cite{Dessart2022A&A...658A.130D}. The comparison interaction configurations are overplotted and normalised to the peak luminosity of SN~2024acyl. }
    \label{fig:24acyl_bollccmp_mdl}
\end{figure}

An alternative scenario for SN\,2024acyl is to invoke the explosion of a low-mass \citep[$M_{\mathrm{preSN}}\lesssim3.5\,M_{\odot}$;][]{Woosley2019ApJ...878...49W, Dessart2022A&A...658A.130D} He star within a binary system, resulting in a partially stripped CC~SN. This binary channel allows the progenitor to be of lower mass, stripped of its H envelope through binary interaction, rather than due to mass loss in a very massive progenitor. As an extreme example, \citet{Sanders2013ApJ...769...39S} explored a white dwarf in a helium-rich CSM within a binary system as a possible progenitor channel for PS1-12sk. This scenario does not require a massive progenitor and recent star formation. 
The derived parameters of SN\,2024acyl seem to well match those of PS1-12sk, such as the low explosion energy ($E_{\mathrm{kin}}$(SN\,2024acyl) = $0.06^{+0.01}_{-0.01} \times 10^{51}$ erg; $E_{\mathrm{kin}}$(PS1-12sk) $\approx 10^{50}$ erg), the small ejecta mass ($M_{\mathrm{ej}}$(SN\,2024acyl) = $0.49^{+0.11}_{-0.09} \, \msun$; $M_{\mathrm{ej}}$(PS1-12sk) $\approx 0.3$\,\msun), along with the presence of H features in the spectra. 
The $M_{\mathrm{Ni}}$ of SN~2024acyl is lower than that of other Type Ibn SNe (which average $\sim$0.04~\msun; \citealt{Maeda2022ApJ...927...25M}) and is comparable to values for low-mass Type IIP SNe \citep{Elmhamdi2003A&A...404.1077E}, suggesting that SN~2024acyl likely originated from a low-mass progenitor.
In addition, SN~2024acyl is located at the edge of the host galaxy, $\sim$34 kpc from its centre, suggesting a potential lower star-formation rate in the nearby area, where it is less likely to form a massive WR-like progenitor \citep{Hosseinzadeh2019ApJL, Warwick2025MNRAS.536.3588W}.
The similarities with PS1-12sk suggest the progenitor of SN~2024acyl is likely a low-mass star, which might even be a helium white dwarf as proposed by \citet{Wang2024MNRAS.530.3906W} for a possible progenitor case of another low-mass event, SN~2020nxt.
Furthermore, the overall photometric and spectroscopic observables of SN\,2024acyl, along with their corresponding modelling results, can be naturally explained by the explosion of a low-mass He star progenitor. 

From a spectroscopic point of view, as discussed in Sec.~\ref{subsec:SpectroModelling}, the non-local thermodynamic equilibrium fits reproduce the main features and favour a low-mass helium-star progenitor --- in particular, a model with a He-ZAMS mass of 4~M$_\odot$, corresponding to a pre-SN mass of 3.16~M$_\odot$.
Building on the interaction-powered framework, the spectra of SN~2024acyl are consistent with emission from a compact CDS formed by the collision of low-mass ejecta with a slowly expanding, He-rich circumstellar shell. In such models, the radiative output is dominated by shock power and non-thermal deposition, with radioactive heating contributing little at the epochs studied \citep{Dessart2022A&A...658A.130D}. The persistence of strong \ion{He}{i} lines, accompanied by weak Balmer features, implies a confined, helium-dominated CSM. These constraints support a low-mass helium-star progenitor, likely stripped via binary interaction, rather than a very massive WR star.
From a photometric point of view (see Sec.~\ref{Sec:MOSFiT}),  the low posteriors of $M_{\mathrm{ej}}$ and $E_{\mathrm{k}}$ suggest SN~2024acyl was a slightly underenergetic event with a small ejected mass. 
In a single-star scenario, the stellar winds from a moderate-mass progenitor are too weak to effectively strip its hydrogen envelope. However, in a close binary system, the progenitor can be stripped through mass transfer to a companion star \citep{Podsiadlowski1992ApJ}.

Finally, we compared the bolometric light curve of SN~2024acyl with a selection of radiation-hydrodynamics models from \cite{Dessart2022A&A...658A.130D} in Fig. \ref{fig:24acyl_bollccmp_mdl}. The bolometric light curve of SN~2024acyl was constructed by fitting black-body models to the SED using an a Markov chain Monte Carlo method, with uncertainties quantified by the 68\% confidence intervals of the posterior distributions. We focused on their simulations of an explosion interacting with a dense, shell-like CSM. These models simulate a low-energy explosion ($E_{\mathrm{k}} = 7.5 \times 10^{49}$ erg) colliding with a $1.0 \,\msun$ CSM shell. The primary comparison is between a high-mass ejecta scenario (models \texttt{E1} and \texttt{E2}, with $M_{\mathrm{ej}} = 1.49 \,\msun$) and a low-mass ejecta scenario (models \texttt{E5} and \texttt{E6}, with $M_{\mathrm{ej}} = 0.15 \,\msun$). 
Within these pairs, the models further differ in the expansion velocity of the CSM, corresponding to CSM kinetic energies of $10^{47}$ erg and $10^{49}$ erg, respectively. The observed light-curve profile of SN~2024acyl shows a strong resemblance to the low-mass ejecta models (\texttt{E5} and \texttt{E6}) rather than high-mass ejecta models (\texttt{E1} and \texttt{E2}). This result suggests that the event was likely produced by an explosion with a small ejecta mass and low kinetic energy, consistent with the light-curve fitting result in Sec.~\ref{Sec:MOSFiT}. Thus, if we have such a small ejecta mass and low kinetic energy, the straightforward interpretation would more likely be that SN~2024acyl was a low-mass stripped helium star.

Furthermore, in the required binary-star scenario, the strong and eruptive mass loss responsible for the dense and shell-like CSM is attributed to late-stage instabilities. Mechanisms such as unstable mass transfer or a nuclear flash can eject a dense, shell-like, helium-rich CSM in the final years before the explosion \citep{Wang2024MNRAS.530.3906W}. This scenario is consistent with our observations, as radiative-hydrodynamics models assuming such a shell-like geometry successfully reproduce the key features of the bolometric light curve of SN~2024acyl.

\section{Conclusions}  
\label{sec:Conclusions}
In this paper, we have presented observations of the recent, well-monitored Type Ibn SN~2024acyl, which has a relatively faint luminosity, linearly and rapidly evolving light curves, and flash-ionisation spectroscopic features. We propose that SN~2024acyl is likely the interaction-powered explosion of a low-mass He star that evolved in an interacting binary, and the CSM with some residual H may have been produced by a mass-transfer process of the binary system. This is based on the observational properties of SN~2024acyl, the multi-band light-curve modelling with \texttt{MOSFiT}, bolometric light curve and spectral comparisons with theoretical models, and a comparison with other SNe Ibn with similar photometric and spectroscopic features. However, a late-type WR star with hydrogen or even an Ofpe/WN9 star with  fallback accretion cannot be entirely ruled out. Lacking a direct detection for the pre-SN mass loss of the progenitor of SN~2024acyl and its companion star, a robust conclusion about the progenitor star, whether arising from a high-mass WR star, a lower-mass He star, or even a He white dwarf in a binary system, cannot be made. 

Advanced observational facilities, such as the Mephisto\footnote{\url{http://www.mephisto.ynu.edu.cn}} and the Vera C. Rubin Observatory,\footnote{\url{https://www.lsst.org/}} will help increase the discovery rate and may trace the progenitor activities of Type Ibn events. New spectroscopic instrumentation such as SOXS \citep[Son Of X-Shooter; see][]{Radhakrishnan2024arXiv240717288R}, installed on the NTT at the La Silla Observatory in Chile, will play a key role in classifying and following up these transients. In addition, the Chinese Space Station Telescope\footnote{\url{http://nao.cas.cn/csst/}}  will enhance our ability to detect and characterise the progenitors of such events. The endeavours we undertake in both theoretical models and advanced facilities will be beneficial to enhancing our understanding of the nature of the poorly understood Type Ibn SNe.

\section{Data availability}
Ultraviolet and optical photometric measurements of SN 2024acyl are available at the Strasbourg astronomical Data Centre via anonymous ftp to cdsarc.cds.unistra.fr (xxx.xx.xxx.x) or via https://cdsarc.cds.unistra.fr/viz-bin/cat/J/A+A/xxx/xxx. Our observations of the spectra are available via the Weizmann Interactive SN Data Repository (WISeREP; \citealt{Yaron2012PASP..124..668Y}).
\bibliographystyle{aa}
\bibliography{Ibnref.bib}
\begin{appendix}

\onecolumn
\section{Observational facilities used in the follow-up photometry of SN\,2024acyl}
\label{appendix:facilities}
\vspace{-1.0em}
\begin{table}[htbp]
\centering
\caption{Information on the instrumental setups.}
\label{table_telescope}
\scalebox{0.8}{
\begin{tabular}{@{}lllll@{}}
\hline \hline
Code & Diameter&Telescope & Instrument & Site \\
                & $\mathrm{m}$&            &                   & \\
\hline
Moravian$^{1}$       & 0.67/0.92 & Schmidt Telescope            & Moravian   &  Osservatorio Astronomico di Asiago, Asiago, Italy\\
TNOT$^{2}$       & 0.8   & Tsinghua-Nanshan Optical Telescope  &   ANDOR  & Nanshan Station of Xinjiang Astronomical Observatory, Xinjiang, P.R. China \\
fa11$^{3}$       & 1.00      & LCO (TFN site)               & Sinistro   & LCO node at Teide Observatory, Tenerife, Spain\\
fa20$^{3}$       & 1.00      & LCO (TFN site)               & Sinistro   & LCO node at Teide Observatory, Tenerife, Spain\\
IO:O$^{4}$           & 2.00      & Liverpool Telescope          & IO:O       &  Observatorio Roque de Los Muchachos, La Palma, Spain\\
ALFOSC$^{5}$         & 2.56      & Nordic Optical Telescope     & ALFOSC     &  Observatorio Roque de Los Muchachos, La Palma, Spain\\
ANDOR$^{6}$          & 1.60      & Multichannel Photometric Survey Telescope & ANDOR & Gaomeigu site, Lijiang Observatory (LJO), Yunnan, P.R. China\\
LJT$^{7}$            & 2.40      & Lijiang 2.4\,m Telescope  & YFOSC      & Gaomeigu site, Lijiang Observatory (LJO), Yunnan, P.R. China\\
ATLAS$^{8}$          & 0.50      & ATLAS                         & ATLAS       & Haleakal\={a} Observatory, Hawaii, USA; Mauna Loa Observatory, Hawaii, USA\\
PS1$^{9}$            & 1.80      & Pan-STARRS1                  & GPC1        & Haleakal\={a} Observatory, Maui, Hawaii, USA \\
UVOT$^{10}$           & 0.30      & Swift Modified Ritchey-Chr\'{e}tien        & UVOT  & Neil Gehrels Swift Observatory \\

\hline \hline
\end{tabular}
}
\vspace{0.3em}
\begin{minipage}{\columnwidth}
\footnotesize
\noindent Sources: 1: A detailed description can be found at \url{https://www.oapd.inaf.it/sede-di-asiago/telescopes-and-instrumentations/schmidt-6792}; 2: Hu et al. 2026, in prep.; 3: \cite{Brown2013PASP..125.1031B}, 4: \cite{Steele2004SPIE.5489..679S}, 5: \cite{Djupvik2010ASSP...14..211D}, 6: \cite{Yuan2020SPIE11445E}, 7: \cite{Wang2019RAA....19..149W}, 8: \cite{Tonry2018PASP..130f4505T}, 9: \cite{Chambers2016arXiv161205560C}, 10: \cite{Roming2005SSRv..120...95R}.
\end{minipage}
\end{table}
\vspace{-2.5em}

\section{Best-fit light-curve model for SN 2024acyl and comparisons}
\label{appendix:LCmodel_parameters}
\vspace{-1.0em}
\begin{table*}[htbp]
    \centering
    \caption{\texttt{MOSFiT} multi-band light-curve fitting results of SN~2024acyl with \texttt{RD+CSI} model. The parameters are converted to linear space.}
    \begin{tabularx}{\columnwidth}{ccYYY}
        \hline
        Parameters& Unit & Prior & Posterior ($n=10$)& Posterior ($n=12$) \\
        \hline
        \multicolumn{5}{c}{\textbf{Shell-like CSM ($s=0$)}} \\
        \hline
        $f_{\mathrm{Ni}}$ & -- &
        $\mathrm{log}~\mathcal{U}~(10^{-3}, 0.1)$ &
        $0.036_{-0.007}^{+0.010}$ &
        $0.063_{-0.016}^{+0.020}$ \\
        
        $E_{\mathrm{k}}$ & $10^{51}\,\mathrm{erg}$ &
        $\mathrm{log}~\mathcal{U}~(0.01, 1.0)$ &
        $0.06_{-0.01}^{+0.01}$ &
        $0.04_{-0.01}^{+0.01}$ \\
        
        $M_{\mathrm{CSM}}$ & $M_\odot$ &
        $\mathrm{log}~\mathcal{U}~(0.05, 0.8)$ &
        $0.51_{-0.04}^{+0.05}$ &
        $0.52_{-0.03}^{+0.06}$ \\
        
        $M_{\mathrm{ej}}$ & $M_\odot$ &
        $\mathrm{log}~\mathcal{U}~(0.1, 10.0)$ &
        $0.49_{-0.09}^{+0.11}$ &
        $0.29_{-0.06}^{+0.10}$ \\
        
        $R_0$ & AU &
        $\mathrm{log}~\mathcal{U}~(10, 30)$ &
        $17.8_{-3.0}^{+3.6}$ &
        $19.5_{-3.7}^{+3.9}$ \\
        
        $\rho$ & $\mathrm{g\,cm^{-3}}$ &
        $\mathrm{log}~\mathcal{U}~(1\times10^{-14}, 8\times10^{-9})$&
        $8.3_{-1.2}^{+2.7}\times 10^{-12}$ &
        $8.9_{-1.5}^{+1.8}\times 10^{-12}$ \\
        \hline \hline
        \multicolumn{5}{c}{\textbf{Wind-like CSM ($s=2$)}} \\
        \hline
        $f_{\mathrm{Ni}}$ & -- &
        $\mathrm{log}~\mathcal{U}~(10^{-3}, 0.1)$ &
        $0.060_{-0.010}^{+0.014}$ &
        $0.085_{-0.016}^{+0.010}$ \\
        
        $E_{\mathrm{k}}$ & $10^{51}\,\mathrm{erg}$ &
        $\mathrm{log}~\mathcal{U}~(0.01, 1.0)$ &
        $0.04_{-0.01}^{+0.01}$ &
        $0.02_{-0.01}^{+0.01}$ \\
        
        $M_{\mathrm{CSM}}$ & $M_\odot$ &
        $\mathrm{log}~\mathcal{U}~(0.05, 0.8)$ &
        $0.54_{-0.04}^{+0.09}$ &
        $0.66_{-0.04}^{+0.05}$ \\
        
        $M_{\mathrm{ej}}$ & $M_\odot$ &
        $\mathrm{log}~\mathcal{U}~(0.1, 10.0)$ &
        $0.30_{-0.05}^{+0.05}$ &
        $0.21_{-0.02}^{+0.05}$ \\
        
        $R_0$ & AU &
        $\mathrm{log}~\mathcal{U}~(10, 30)$ &
        $16.6_{-0.8}^{+1.2}$ &
        $14.1_{-2.6}^{+3.3}$ \\
        
        $\rho$ & $\mathrm{g\,cm^{-3}}$ &
        $\mathrm{log}~\mathcal{U}~(1\times10^{-14}, 8\times10^{-9})$ &
        $9.8_{-2.5}^{+1.5}\times 10^{-12}$ &
        $8.7_{-2.4}^{+3.6}\times 10^{-12}$ \\
        
        \hline
    \end{tabularx}
    \label{apptab:lcmodeling}
    \vspace{0.4em}
    \scriptsize 
    \raggedright
    
    \textbf{Notes:} $\mathrm{log}~\mathcal{U}(a, b)$ denotes a Log-Uniform prior distribution between $a$ and $b$. $\mathcal{U}(a, b)$ denotes a Uniform distribution. Posterior values for log-parameters have been converted to a linear scale (median with $16^{\mathrm{th}}$ and $84^{\mathrm{th}}$ percentiles mapped accordingly).
\end{table*}
\vspace{-2.5em}

\section{Log of spectroscopic observations}
\label{Spec_log}
\vspace{-1.0em}
\begin{table}[htbp]
\centering
\caption{Log of the spectroscopic observations of SN~2024acyl.}
\label{table:speclog_2024acyl}
\small
\setlength{\tabcolsep}{5pt}
\begin{tabular}{cccccccc}
\hline\hline
Date & MJD & Phase$^{a}$ & Instrumental setup & Grism/Grating & Spectral range & Exposure time & Resolution \\
     &     & (d) & & & (\AA) & (s) & (\AA) \\
\hline
20241204 & 60648.8 & $-3.7$ & 0.3\,m Telescope+N300 & ALPY200 & 3800--7400 & 1800$\times$4 & $R \approx 130$ \\
20241206 & 60650.1 & $-2.4$ & NTT+EFOSC2 & gr13 & 3280--9270 & 1500 & 21 \\
20241207 & 60651.7 & $-0.8$ & LJT+YFOSC & gm3 & 3610--8900 & 2200 & 18 \\
20241208 & 60652.1 & $-0.4$ & NTT+EFOSC2 & gr11 & 3650--9240 & 2300 & 16 \\
20241209 & 60653.0 & $+0.5$ & TNG+DOLORES & LR-B & 3340--8120 & 1500 & 15 \\
20241209 & 60653.0 & $+0.5$ & TNG+DOLORES & VHR-V & 5070--6780 & 1500 & 3.5 \\
20241209 & 60653.2 & $+0.7$ & NTT+EFOSC2 & gr11 & 3380--7480 & 3600 & 16 \\
20241210 & 60654.0 & $+1.5$ & NTT+EFOSC2 & gr11 & 3370--7490 & 3600 & 16 \\
20241210 & 60654.3 & $+1.8$ & Lick Shane+Kast & 4310/7500 & 3610--10,750 & 2760/2700 & 9.3 \\
20241224 & 60668.1 & $+15.6$ & NTT+EFOSC2 & gr11 & 3370--7480 & 3600 & 16 \\
20241227 & 60671.1 & $+18.6$ & NTT+EFOSC2 & gr16 & 6020--10,000 & 3600 & 16 \\
20250107 & 60682.1 & $+29.6$ & NTT+EFOSC2 & gr11 & 3370--7480 & 3600 & 16 \\
20250120 & 60695.3 & $+42.8$ & Gemini North+GMOS-N & B480 & 3650--9240 & 900$\times$2+900$\times$2 & $R\approx 750$ \\
\hline\hline
\end{tabular}
\vspace{0.3em}
\begin{minipage}{\linewidth}
\footnotesize
$^{a}$Phases are relative to $o$-band maximum light (MJD = $60652.49 \pm 0.26$; 2024-12-08) in the observer frame.
\end{minipage}
\end{table}
\vspace{-2.5em}

\newpage
\section{MOSFiT corner plots of SN\,2024acyl} \label{appendix:corner}
\begin{figure}[htb]
    \centering
    \includegraphics[width=1\linewidth]{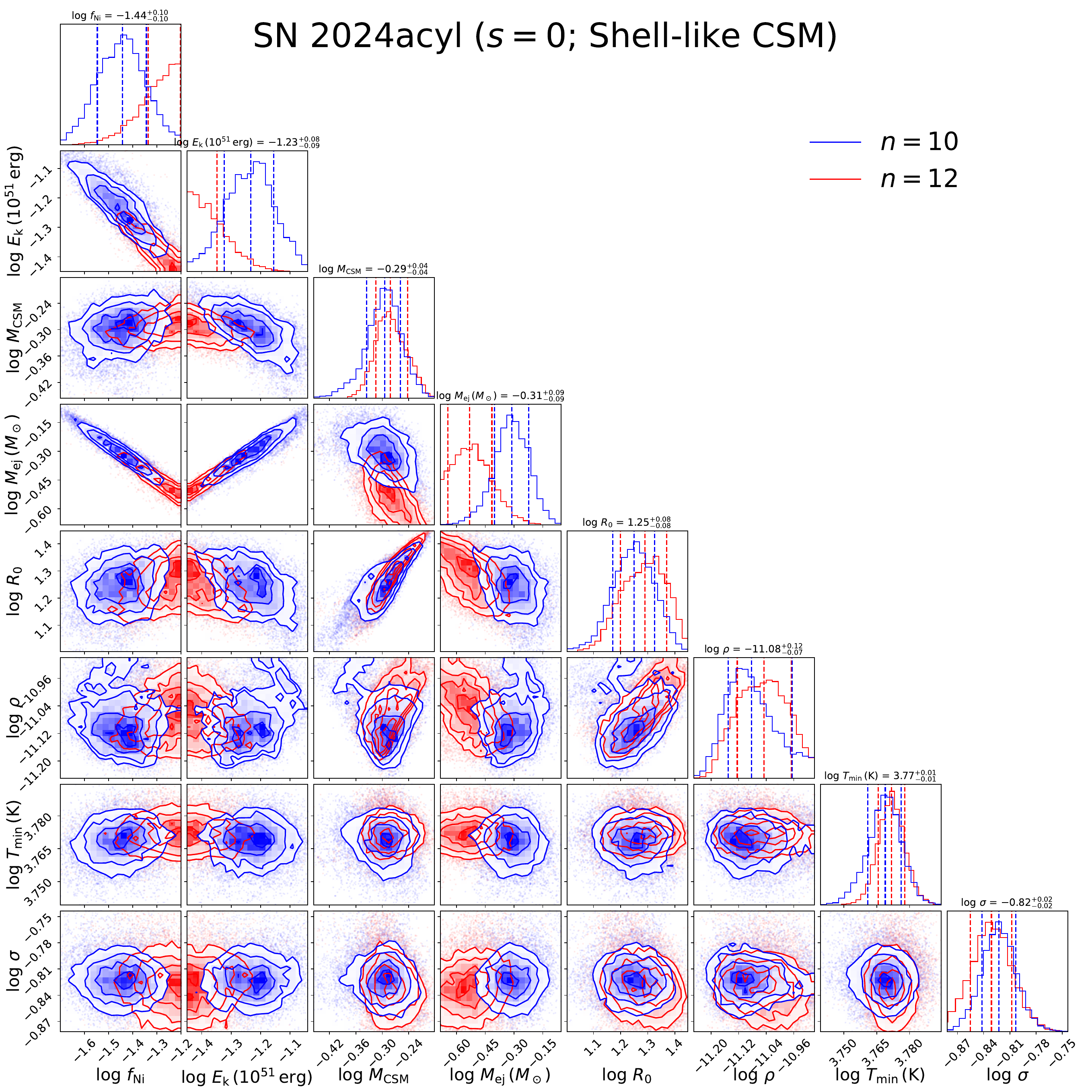}
    \caption{Corner plots showing the posterior distributions of the estimated parameters for SN\,2024acyl, based on the \texttt{Ni+CSM} model using \texttt{MOSFiT}. All the models adopted the CSM density profile $s=0$ (shell-like CSM). The blue region represents outer ejecta density profile $n=10$ while red region represents $n=12$. Median values are marked by  dashed vertical lines, with the shaded regions representing the 68\% confidence intervals.}
    \label{fig:modfit_corner1}
\end{figure}

\section{Acknowledgements}
{\tiny
We are grateful to the anonymous referee for insightful comments and suggestions that improved the paper.
We thank Luc Dessart for kindly providing the spectral models for this Type Ibn SN. 
This work is supported by the National Key Research and Development Program of China (grants 2024YFA1611603, 2021YFA1600404), the National Natural Science Foundation of China (grants 12303054, 12225304, 12288102, 12373038), the Yunnan Fundamental Research Projects (grants 202401AU070063, 202501AS070078), the Yunnan Revitalization Talent Support Program (Yunling Scholar Project and Innovation Team Project), the Yunnan Science and Technology Program (grants 202501AS070005, 202201BC070003), the International Centre of Supernovae, Yunnan Key Laboratory (grant 202302AN360001), the Natural Science Foundation of Xinjiang Uygur Autonomous Region (grant 2024D01D32), and the Tianshan Talent Training Program (grant  2023TSYCLJ0053).
A.P., A.R., S.B., E.C., N.E.R., and G.V. acknowledge support from the PRIN-INAF 2022, ``Shedding light on the nature of gap transients: from the observations to the models.''
A.R. was also supported by the GRAWITA Large Program Grant (PI P. D'Avanzo).
E.C. acknowledges support from MIUR, PRIN 2020 (METE, grant 2020KB33TP).
K.M. was supported by Japan Society for the Promotion of Science (JSPS) KAKENHI grants  JP24KK0070 and JP24H01810, and by JSPS bilateral program JPJSBP120229923. 
B.K. is supported by the ``Special Project for High-End Foreign Experts,'' Xingdian Funding from Yunnan Province.
T.-W.C. acknowledges financial support from the Yushan Fellow Program by the Ministry of Education, Taiwan (MOE-111-YSFMS-0008-001-P1) and the National Science and Technology Council, Taiwan (NSTC grant 114-2112-M-008-021-MY3).
N.E.R. acknowledges support from the Spanish Ministerio de Ciencia e Innovaci\'on (MCIN) and the Agencia Estatal de Investigaci\'on (AEI) 10.13039/501100011033 under the program Unidad de Excelencia Mar\'ia de Maeztu CEX2020-001058-M.
C.P.G. acknowledges financial support from the Secretary of Universities and Research (Government of Catalonia) and by the Horizon 2020 Research and Innovation Programme of the European Union under the Marie Sk\l{}odowska-Curie and the Beatriu de Pin\'os 2021 BP~00168 programme. 
Both C.P.G. and M.G.B. were supported by the Spanish Ministerio de Ciencia e Innovaci\'on (MCIN) and the Agencia Estatal de Investigaci\'on (AEI; 10.13039/501100011033) under the PID2023-151307NB-I00 SNNEXT project, by the Centro Superior de Investigaciones Cient\'ificas (CSIC) under the PIE project 20215AT016 and the programme Unidad de Excelencia Mar\'ia de Maeztu CEX2020-001058-M, and by the Departament de Recerca i Universitats de la Generalitat de Catalunya through the 2021-SGR-01270 grant.
T.K. acknowledges support from the Research Council of Finland project 360274.
S. Mattila acknowledges financial support from the Research Council of Finland project 350458.
S. Moran is funded by Leverhulme Trust grant RPG-2023-240.
M.D. Stritzinger is funded by the Independent Research Fund Denmark (IRFD, grant number 10.46540/2032-00022B).
T.E.M.B. is funded by Horizon Europe ERC grant no. 101125877.
T.P. acknowledges financial support from the Slovenian Research Agency (grants I0-0033, P1-0031, J1-2460 and N1-0344).
H.K. was funded by the Research Council of Finland projects 324504, 328898, and 353019.
A.F. acknowledges funding by the European Union -- NextGenerationEU RFF M4C2 1.1 PRIN 2022 project ``2022RJLWHN URKA'' and by INAF 2023 Theory Grant ObFu 1.05.23.06.06 ``Understanding R-process \& Kilonovae Aspects (URKA).''
J.Z. is supported by the National Key R\&D Program of China with No. 2021YFA1600404, the National Natural Science Foundation of China (12173082, 12333008), the Yunnan Fundamental Research Projects (grants 202401BC070007 and 202201AT070069), the Top-notch Young Talents Program of Yunnan Province, the Light of West China Program provided by the Chinese Academy of Sciences, and the International Centre of Supernovae, Yunnan Key Laboratory (grant 202302AN360001).
X.F.W. is supported by the National Natural Science Foundation of China (grants 12288102, 12033003, and 11633002) and the Tencent Xplorer Prize.
A.V.F.'s research group at U.C. Berkeley acknowledges financial assistance from the Christopher R. Redlich Fund, as well as donations 
from Gary and Cynthia Bengier, Clark and Sharon Winslow, Alan Eustace and Kathy Kwan, William Draper, Timothy and Melissa Draper, Briggs and Kathleen Wood, Sanford Robertson (W.Z. is a Bengier-Winslow-Eustace Specialist in Astronomy, T.G.B. is a Draper-Wood-Robertson Specialist in Astronomy), and numerous other donors.    

We acknowledge the support of the staffs of the various observatories at which data were obtained.
The observations were in part carried out within the framework of Subaru-Gemini time exchange program (under the project S24B-041 / GN-2024B-Q-101: PI, K. Maeda) which is operated by the National Astronomical Observatory of Japan. We are honored and grateful for the opportunity of observing the Universe from Maunakea, which has the cultural, historical and natural significance in Hawaii.
Funding for the LJT has been provided by the CAS and the People’s Government of Yunnan Province. The LJT is jointly operated and administrated by YNAO and the Center for Astronomical Mega-Science, CAS.
Based in part on observations made with the Nordic Optical Telescope, owned in collaboration by the University of Turku and Aarhus University, and operated jointly by Aarhus University, the University of Turku, and the University of Oslo, representing Denmark, Finland, and Norway, the University of Iceland, and Stockholm University at the Observatorio del Roque de los Muchachos, La Palma, Spain, of the Instituto de Astrofisica de Canarias.
Observations from the NOT were obtained through the NUTS2 collaboration which is supported in part by the Instrument Centre for Danish Astrophysics (IDA), and the Finnish Centre for Astronomy with ESO (FINCA) via Academy of Finland grant nr 306531. The data presented here were obtained in part with ALFOSC, which is provided by the Instituto de Astrofisica de Andalucia (IAA) under a joint agreement with the University of Copenhagen and NOTSA.
The Liverpool Telescope is operated on the island of La Palma by Liverpool John Moores University in the Spanish Observatorio del Roque de los Muchachos of the Instituto de Astrofisica de Canarias with financial support from the UK Science and Technology Facilities Council.
The Italian Telescopio Nazionale Galileo (TNG) operated on the island of La Palma by the Fundaci\'on Galileo Galilei of the INAF (Istituto Nazionale di Astrofisica) at the Spanish Observatorio del Roque de los Muchachos of the Instituto de Astrofísica de Canarias. This article is also based on observations made in the Observatorios de Canarias del IAC with the Telescopio Nazionale Galileo, operated on the island of La Palma by INAF at the Observatorio del Roque de los Muchachos under the program A50TAC\_41 (PI: G. Valerin).
Based in part on observations collected at Copernico and Schmidt telescopes (Asiago Mount Ekar, Italy) of the INAF -- Osservatorio Astronomico di Padova.
The Chinese Tsinghua–Nanshan Optical Telescope (TNOT) operated at Nanshan Station by Xinjiang Astronomical Observatory of the Chinese Academy of Sciences, located in Xinjiang, China.
Based in part on observations collected with the 0.8\,m TNOT equipped with an Andor camera at Nanshan Station of Xinjiang Astronomical Observatory. 
Mephisto is developed at and operated by the South-Western Institute for Astronomy Research of Yunnan University (SWIFAR-YNU), funded by the ``Yunnan University Development Plan for World-Class University'' and ``Yunnan University Development Plan for World-Class Astronomy Discipline.'' 
Based on the “Key Laboratory of Survey Science of Yunnan Province” with project No. 202449CE340002.
Based in part on observations collected at the European Organisation for Astronomical Research in the Southern Hemisphere, Chile, as part of ePESSTO+ (the advanced Public ESO Spectroscopic Survey for Transient Objects Survey – PI: Inserra). ePESSTO+ observations were obtained under ESO program ID 112.25JQ.
A major upgrade of the Kast spectrograph on the Shane 3\,m telescope at        
Lick Observatory, led by Brad Holden, was made possible through generous gifts from the Heising-Simons Foundation, William and Marina Kast, and the University of California Observatories.  Research at Lick Observatory is partially supported by a generous gift from Google. 

This work has made use of data from the Asteroid Terrestrial-impact Last Alert System (ATLAS) project. The Asteroid Terrestrial-impact Last Alert System (ATLAS) project is primarily funded to search for near-Earth objects (NEOs) through National Aeronautics and Space Administration (NASA) grants NN12AR55G, 80NSSC18K0284, and 80NSSC18K1575; byproducts of the NEO search include images and catalogs from the survey area. This work was partially funded by Kepler/K2 grant J1944/80NSSC19K0112 and HST GO-15889, and STFC grants ST/T000198/1 and ST/S006109/1. The ATLAS science products have been made possible through the contributions of the University of Hawaii Institute for Astronomy, the Queen's University Belfast, the Space Telescope Science Institute, the South African Astronomical Observatory, and The Millennium Institute of Astrophysics (MAS), Chile.

Pan-STARRS is a project of the Institute for Astronomy of the University of Hawaii, and is supported by the NASA SSO Near Earth Observation Program under grants 80NSSC18K0971, NNX14AM74G, NNX12AR65G, NNX13AQ47G, NNX08AR22G, 80NSSC21K1572 and by the State of Hawaii. The Pan-STARRS1 Surveys (PS1) and the PS1 public science archive have been made possible through contributions by the Institute for Astronomy, the University of Hawaii, the Pan-STARRS Project Office, the Max-Planck Society and its participating institutes, the Max Planck Institute for Astronomy, Heidelberg and the Max Planck Institute for Extraterrestrial Physics, Garching, The Johns Hopkins University, Durham University, the University of Edinburgh, the Queen's University Belfast, the Harvard-Smithsonian Center for Astrophysics, the Las Cumbres Observatory Global Telescope Network Incorporated, the National Central University of Taiwan, STScI, NASA under grant NNX08AR22G issued through the Planetary Science Division of the NASA Science Mission Directorate, NSF grant AST-1238877, the University of Maryland, Eotvos Lorand University (ELTE), the Los Alamos National Laboratory, and the Gordon and Betty Moore Foundation. 

We acknowledge the use of public data from the \textit{Swift} data archive. SDSS is managed by the Astrophysical Research Consortium for the Participating Institutions of the SDSS Collaboration including the Brazilian Participation Group, the Carnegie Institution for Science, Carnegie Mellon University, Center for Astrophysics \textbar\ Harvard \& Smithsonian (CfA), the Chilean Participation Group, the French Participation Group, Instituto de Astrof\'isica de Canarias, The Johns Hopkins University, Kavli Institute for the Physics and Mathematics of the Universe (IPMU) / University of Tokyo, the Korean Participation Group, Lawrence Berkeley National Laboratory, Leibniz Institut f\"{u}r Astrophysik Potsdam (AIP), Max-Planck-Institut f\"{u}r Astronomie (MPIA Heidelberg), Max-Planck-Institut f\"{u}r Astrophysik (MPA Garching), Max-Planck-Institut f\"{u}r Extraterrestrische Physik (MPE), National Astronomical Observatories of China, New Mexico State University, New York University, University of Notre Dame, Observat\'orio Nacional / MCTI, The Ohio State University, Pennsylvania State University, Shanghai Astronomical Observatory, United Kingdom Participation Group, Universidad Nacional Autónoma de México, University of Arizona, University of Colorado Boulder, University of Oxford, University of Portsmouth, University of Utah, University of Virginia, University of Washington, University of Wisconsin, Vanderbilt University, and Yale University. This research has made use of the NASA/IPAC Extragalactic Database (NED), which is operated by the Jet Propulsion Laboratory, California Institute of Technology, under contract with NASA. \par
}
\end{appendix}
\end{document}